\documentclass[prb,amsfonts,twocolumn,showpacs,amssymb,floatfix,bibnotes]{revtex4}
\usepackage{bm}
\usepackage{tabularx}
\usepackage{ascmac}
\usepackage{amsmath}
\usepackage{wrapfig}
\usepackage{graphicx}
\usepackage{float}
\usepackage[FIGBOTCAP]{subfigure}
\usepackage{color}

\bmdefine{\bolds}{s}
\bmdefine{\boldi}{i}
\bmdefine{\boldj}{j}
\bmdefine{\boldiprime}{i^{\prime}}
\bmdefine{\boldjprime}{j^{\prime}}
\bmdefine{\boldtau}{\tau}
\bmdefine{\boldsigma}{\sigma}
\bmdefine{\boldk}{k}
\bmdefine{\boldK}{K}
\bmdefine{\boldr}{r}

\begin{document}


\title{
Origin of the Heavy Fermion Behavior in 
Ca$_{2-x}$Sr$_{x}$RuO$_{4}$:\\ Roles 
of Coulomb Interaction and the Rotation of RuO$_{6}$ octahedra }


\author{Naoya Arakawa}
\email{arakawa@hosi.phys.s.u-tokyo.ac.jp} 
\author{Masao Ogata}
\affiliation{Department of Physics, 
The University of Tokyo,
Tokyo 113-0033, Japan}


\date{\today}

\begin{abstract}
We study the electronic states for Ca$_{2-x}$Sr$_{x}$RuO$_{4}$ 
in $0.5\leq x \leq 2$ within the Gutzwiller approximation (GA) 
on the basis of the three-orbital Hubbard model for the Ru $t_{2g}$ orbitals. 
The main effects of the Ca-substitution 
are taken into account as the changes of the $dp$ hybridizations 
between the Ru $4d$ and O $2p$ orbitals. 
Using the numerical minimization of the energy obtained in the GA, 
we obtain the renormalization factor (RF) of the kinetic energy 
and total RF, which estimates the inverse of the mass enhancement, 
for three cases with the effective models 
of $x=2$ and $0.5$ and a special model. 
We find that 
the inverse of the total RF 
becomes the largest for the case of $x=0.5$, 
and that 
the van Hove singularity, 
which is located on (below) the Fermi level for 
the special model (the effective model of $x=0.5$), plays a secondary role 
in enhancing the effective mass. 
Our calculation suggests that 
the heavy fermion behavior around $x=0.5$ comes from 
the cooperative effects between 
moderately strong Coulomb interaction compared to 
the total bandwidth 
and the modification of the electronic structures 
due to the rotation of RuO$_{6}$ octahedra (i.e., 
the variation of the $dp\pi$ hybridizations 
and the downward shift for the $d_{xy}$ orbital). 
We propose that 
moderately strong electron correlation and 
the orbital-dependent modifications of the electronic structures 
due to the lattice distortions 
play important roles in the electronic states for Ca$_{2-x}$Sr$_{x}$RuO$_{4}$. 
\end{abstract}


\maketitle

\section{Introduction}
Strongly correlated electron systems with 
orbital degrees of freedom have attracted much interests 
due to a diversity of phenomena, e.g., a structure-sensitive 
superconducting transition temperature,~\cite{Lee} 
a colossal magneto-resistance,~\cite{GMR-Mn-1,GMR-Mn-2} and 
an orbital ordering and 
the corresponding collective excitations.~\cite{OO-1,OO-2} 
Among them, 
the layered perovskite Ca$_{2-x}$Sr$_{x}$RuO$_{4}$ 
has rich ground states,~\cite{Nakatsuji-discovery,
Nakatsuji-lattice}  
in which the Ru $4d$ orbitals play 
important roles in determining the electronic states. 

\begin{figure}[tb]
\includegraphics[width=82mm]{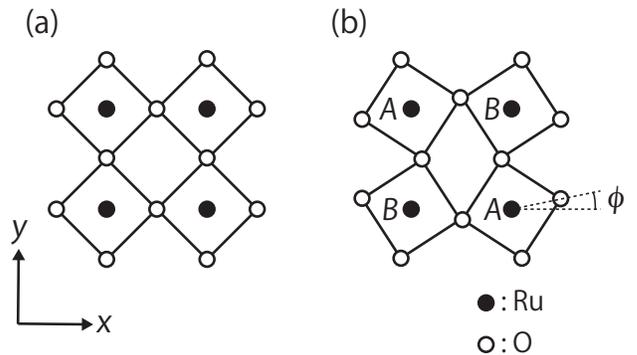}
\vspace{-6pt}
\caption{Schematic pictures of a $xy$ plane of RuO$_{6}$ octahedra 
for (a) $1.5\leq x\leq 2$ and (b) $0.5\leq x< 1.5$. 
Black (white) circles represent Ru (O) ions. 
$\phi$ is the angle of the rotation of RuO$_{6}$ octahedra. 
$A$ and $B$ are the indices of two sublattices.}
\label{fig:rotation}
\end{figure}

In Ca$_{2-x}$Sr$_{x}$RuO$_{4}$, 
substitution of Ca$^{2+}$ for Sr$^{2+}$ causes 
three kinds of lattice distortions, i.e., rotation, tilting, and flattening, 
which affect the electronic structures.~\cite{Nakatsuji-lattice,Lattice} 
Starting from $x=2$ (i.e., Sr$_{2}$RuO$_{4}$), which is 
a spin-triplet superconductor,~\cite{Maeno-triplet,Ishida-NMR,neutron-triplet} 
the ground state changes from the spin-triplet superconductor 
to a paramagnetic (PM) metal in the range of $0.5\leq x < 2$. 
In particular, 
in the range of $0.5\leq x < 1.5$, 
the Ca substitution induces a rotation of RuO$_{6}$ octahedra 
around the $c$ axis (see Fig. \ref{fig:rotation}), 
while the crystalline structure remains tetragonal 
with a unit cell doubled. 
Except the vicinity of $x=2$, 
the spin susceptibility behaves like 
the Curie-Weiss type with a Curie constant 
corresponding to nearly $S=1/2$, and 
the Weiss temperature reaches nearly zero 
at $x=0.5$.~\cite{Nakatsuji-discovery} 
In the range of $0.2\leq x < 0.5$, 
there occurs a structural transition 
from tetragonal phase to orthorhombic phase 
at low temperatures, and 
the Ca substitution induces the tilting of RuO$_{6}$ octahedra 
around a non-symmetric axis in the basal plane. 
Also, in this range, 
the spin susceptibility has a peak at some temperature, and 
the ground state becomes an antiferromagnetically 
correlated metal.~\cite{Nakatsuji-lattice} 
In the range of $0\leq x < 0.2$, 
a flattening of RuO$_{6}$ octahedra along the $c$ axis is induced 
and there is a first-order metal-insulator transition: 
the ground state becomes 
the antiferromagnetic insulator.~\cite{Nakatsuji-Ca}    

Most surprisingly, 
the coefficient of the electronic specific heat, $\gamma_{\textrm{e}}$, 
at low temperatures becomes large around $x=0.5$. 
The largest value of $\gamma_{\textrm{e}}$ reaches 
$255$ mJ/mol-Ru K$^{2}$ at $x=0.5$; 
this value is about $25$ times of that 
obtained in the density-functional calculation for Sr$_{2}$RuO$_{4}$ 
within local-density approximation (LDA).~\cite{Nakatsuji-HF} 
A similar mass enhancement has been observed 
in the optical conductivity measurement 
by using the extended Drude model.~\cite{opt-mass} 
In addition, 
the spin susceptibility and Hall coefficient measured 
by applying the magnetic field perpendicular to $ab$ plane 
show peak structures at $T^{\ast}\sim$ $10$ K for $x=0.3$,~\cite{HF-03} 
which are similar to those obtained for UPt$_{3}$.~\cite{Hall-UPt3}
These experimental results indicate 
a formation of heavy fermions (HFs) around $x=0.5$. 

Although there have been extensive theoretical works, 
the origin of this HF behavior has not been clarified yet. 
Some studies~\cite{Anisimov,Koga} based on 
the dynamical mean-field theory (DMFT) have proposed that 
the HF behavior results from the orbital-selective Mott transition (OSMT) 
for the $d_{xz}$ and $d_{yz}$ orbitals; 
the OSMT is defined as 
a partial disappearance of quasiparticles 
only for some of the conducting bands. 
However, 
this contradicts with another DMFT calculation~\cite{Liebsch} 
which has used a realistic model of 
Ca$_{2-x}$Sr$_{x}$RuO$_{4}$ in $0.5\leq x \leq 2$: 
the OSMT does not appear. 

The OSMT for the $d_{xz}$ and $d_{yz}$ orbitals 
also contradicts with several experiments 
in Ca$_{2-x}$Sr$_{x}$RuO$_{4}$ around $x=0.5$. 
The angle-resolved photoemission spectroscopy (ARPES) 
measurement~\cite{ARPES05} for $x=0.5$ and $2$ has shown that 
the Ca substitution does not modify the topologies of 
the Fermi surfaces (FSs) for the $d_{xz}$ and $d_{yz}$ orbitals. 
This result is inconsistent with the proposal of the OSMT 
since the partial Mott gap should drastically affect the topologies of the FSs. 
In addition, 
the optical conductivity measurement~\cite{opt-mass} 
for Ca$_{2-x}$Sr$_{x}$RuO$_{4}$ in $0.06\leq x \leq 2$ has found that 
the Drude weights depend weakly on the Ca concentration. 
This is consistent with the ARPES measurement 
since the latter shows a small change of the FS 
from that for Sr$_{2}$RuO$_{4}$. 
Note that the FS for the $d_{xy}$ orbital changes 
from an electron pocket for $x=2$ to a hole pocket for $x=0.5$. 

The aim of this paper is 
to discuss the qualitative origin of the HF behavior 
in particular near $x=0.5$ for Ca$_{2-x}$Sr$_{x}$RuO$_{4}$. 
We study the electronic states for $0.5\leq x \leq 2$ 
within the Gutzwiller approximation (GA) 
on the basis of the three-orbital Hubbard model 
for the Ru $t_{2g}$ orbitals (i.e., $d_{xz}$, $d_{yz}$, and $d_{xy}$ orbitals). 
We assume that 
the Ca substitution affects the electronic structures 
mainly by the changes of the $dp$ hybridizations 
between the Ru $4d$ and O $2p$ orbitals. 
Actually,  
the density-functional calculation within the LDA has found that 
the rotation of RuO$_{6}$ octahedra affects 
the electronic structures for Ca$_{2-x}$Sr$_{x}$RuO$_{4}$ 
in $0.5\leq x \leq 2$.~\cite{Terakura} 
The GA is used to include the effects of electron correlation 
non-perturbatively, in which 
the effects lead to the renormalization of the kinetic energy.~\cite{GA,
GA-Ogawa,GA-Fazekas,GA-multi,GA-tJ} 

Using the numerical minimization of the energy obtained in the GA, 
we obtain the renormalization factor (RF) of the kinetic energy 
for the Ru $t_{2g}$ orbtials 
and total RF, which estimates the inverse of the mass enhancement, 
for the effective models of $x=2$ and $0.5$. 
We find that the difference between the total RFs 
for the cases of $x=2$ and $0.5$ becomes large 
as the intraorbital Coulomb interaction is strong; 
this arises from the criticality approaching the usual Mott transition, 
where the occupation numbers for the $d_{xz/yz}$ 
and $d_{xy}$ orbitals are $1$ and $2$, respectively.  
In addition, we analyze a special model in which 
the van Hove singularity (vHs) for the $d_{xy}$ orbital 
is located on the Fermi level. 
We find that 
the total RF is smaller for the effective model of $x=0.5$ 
than that for the special model; 
the vHs plays a secondary role in enhancing the effective mass. 
These results are consistent with the experimentally observed tendency 
of the effective mass in $0.5\leq x \leq 2$, i.e., 
monotonic increase of $\gamma_{\textrm{e}}$ towards $x=0.5$.~\cite{Nakatsuji-HF} 
Our calculation suggests that 
the HF behavior around $x=0.5$ comes from 
the cooperative effects between 
moderately strong Coulomb interaction compared with 
the total bandwidth and the significant modification 
of the electronic structures for the Ru $t_{2g}$ orbitals 
due to the rotation of RuO$_{6}$ octahedra; 
the latter includes both 
the variation of the $dp\pi$ hybridizations and 
the downward shift for the $d_{xy}$ orbital. 

The paper is organized as follows. 
Section II is devoted to the explanations 
of the method to take account 
of the main effects of the Ca substitution 
on the electronic structures in $0.5\leq x\leq 2$ 
and the GA for a PM state of 
the degenerate $d_{xz}$ and $d_{yz}$ orbitals and the $d_{xy}$ orbital. 
In Sec. III, we show the numerical results of the GA 
for three cases with the effective models of $x=2$ and $0.5$ 
and the special model. 
In Sec. IV, 
we compare our results with previous theoretical studies 
and remark on the correspondence of our results with experimental results. 
The paper concludes with a summary of our results in Sec. V. 

\section{Formulation}
\begin{figure}[tb]
\vspace{6pt}
\includegraphics[width=86mm]{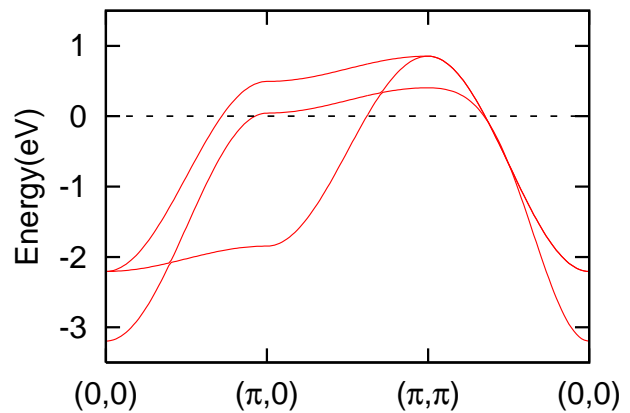}
\vspace{-20pt}
\caption{(Color online) Band structure of the Ru $t_{2g}$ orbitals 
for the effective model of $x=2$. 
The dashed black line represents 
the chemical potential.}
\label{fig:band}
\end{figure}
\begin{figure}[tb]
\vspace{-36pt}
\hspace{-50pt}
\includegraphics[width=102mm]{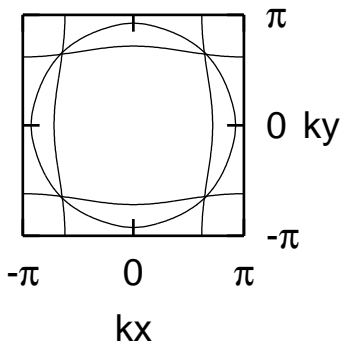}
\vspace{-30pt}
\caption{FSs for the effective model of $x=2$.}
\label{fig:FS}
\end{figure}
\begin{figure}[tb]
\vspace{6pt}
\includegraphics[width=88mm]{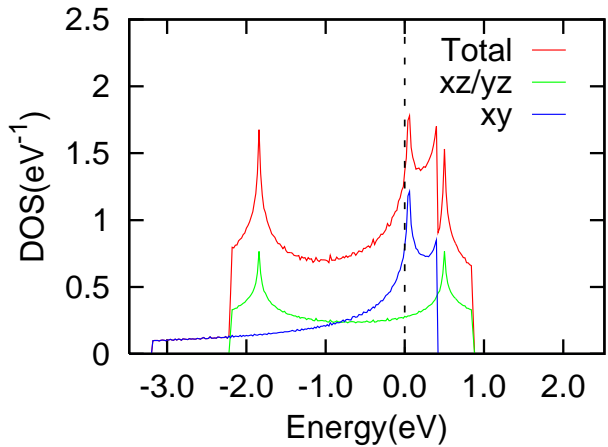}
\vspace{-16pt}
\caption{(Color online) DOS for the effective model of $x=2$. 
The dashed black line represents 
the chemical potential.}
\label{fig:DOS}
\end{figure}
In the following, 
we choose the coordinates, $x$, $y$, and $z$, 
in the directions of the bonds between Ru and O ions 
at $\phi=$ 0$^{\circ}$, 
where $\phi$ is an angle of the rotation of 
RuO$_{6}$ octahedra [see Fig. \ref{fig:rotation}(b)]. 
Namely, 
the coordinates are fixed in the directions of Sr$_{2}$RuO$_{4}$ 
with $\phi=$ 0$^{\circ}$ even for the cases with finite values of $\phi$. 
For convenience, 
the five Ru $4d$ orbitals, $d_{xz}$, $d_{yz}$, $d_{xy}$, 
$d_{x^{2}\textrm{-}y^{2}}$, and $d_{3z^{2}\textrm{-}r^{2}}$, 
are labeled $1$, $2$, $3$, $4$, and $5$, respectively. 

According to the density-functional calculations~\cite{Oguchi,Singh} 
for Sr$_{2}$RuO$_{4}$ within the LDA, 
the antibonding bands of the Ru $t_{2g}$ and O $2p$ orbitals 
form the conducting bands in the vicinity of the Fermi level, and  
the density of states (DOS) near the Fermi level 
is originated mainly from the Ru $t_{2g}$ orbitals. 
The partial density of states (pDOS) for the O $2p$ orbitals 
is roughly a quarter of those for the Ru $t_{2g}$ orbitals. 
In addition, 
the x-ray measurements in Ca$_{2-x}$Sr$_{x}$RuO$_{4}$ 
have shown that 
the crystalline-electric-field (CEF) energy 
between the Ru $e_{g}$ orbitals and Ru $t_{2g}$ orbitals is 
of the order of 1 eV.~\cite{X-ray10Dq,X-ray10Dq-family}  
Thus, the Ru $t_{2g}$ orbitals play main roles in determining 
the electronic states for Ca$_{2-x}$Sr$_{x}$RuO$_{4}$ at low temperatures. 
In Sec. IV, we remark on the roles of the Ru $e_{g}$ and O $2p$ orbitals. 

In order to study the electronic states, 
we use the three-orbital Hubbard model for the Ru $t_{2g}$ orbitals. 
The unit cell contains one Ru atom for $1.5\leq x \leq 2$, 
while it contains two Ru atoms for $0.5\leq x <1.5$
due to the rotation of RuO$_{6}$ octahedra. 
As shown in Fig. \ref{fig:rotation} (b), 
the directions of the rotation alternate in the two-dimensional 
square lattice, leading to a unit cell doubled. 

The noninteracting Hamiltonian is 
\begin{align}
\hat{H}_{0}
=& - 
\textstyle\sum\limits
_{\boldi, \boldj}
\textstyle\sum\limits_{a,b=1}^{3}
\textstyle\sum\limits_{\sigma}
( t_{ab}^{\boldi, \boldj}(\phi) 
\hat{c}^{\dagger}_{\boldi a \sigma} 
\hat{c}_{\boldj b \sigma} 
+ \textrm{H.c.} ) 
-\mu \textstyle\sum\limits_{\boldi}
\textstyle\sum\limits_{a=1}^{3} \hat{n}_{\boldi a} , \label{eq:H0}
\end{align}
where $\hat{c}^{\dagger}_{\boldi a \sigma}$ 
($\hat{c}_{\boldi a \sigma}$) 
is the creation (annihilation) operator that 
creates (annihilates) an electron in $t_{2g}$ orbital $a$($=1,2,3$)
with spin $\sigma$($= \uparrow, \downarrow)$ at site $\boldi$, 
$\hat{n}_{\boldi a}=\sum_{\sigma}\hat{n}_{\boldi a \sigma}
=\sum_{\sigma}\hat{c}^{\dagger}_{\boldi a \sigma} \hat{c}_{\boldi a \sigma}$, 
and H.c. means the Hermitian conjugate.  
Here, $t_{ab}^{\boldi, \boldj}(\phi)$ and $\mu$ denote 
the values of in-plane hopping integrals at an angle $\phi$ 
and the chemical potential, respectively. 
$\mu$ is determined so as to satisfy $n_{\textrm{e}}=4$ with
$n_{\textrm{e}}$ being the total occupation number for the Ru $t_{2g}$ orbitals. 
For simplicity, we neglect the effect of the spin-orbit interaction 
on the electronic structures; the validity is addressed in Sec. IV. 

The hopping integrals for Sr$_{2}$RuO$_{4}$ with $\phi=0^{\circ}$ 
are determined~\cite{Yanase} so as to reproduce the FSs obtained in the 
de Haas-van Alphen effect:~\cite{dHvA} 
the dispersions are given by
\begin{align}
\epsilon_{11}(\boldk, 0^{\circ})&
=-2 t_{1} \cos k_{x}
-2 t_{2} \cos k_{y} -\mu, \label{eq:dis11-0}\\
\epsilon_{22}(\boldk, 0^{\circ})&
=-2t_{2} \cos k_{x}
-2t_{1} \cos k_{y} -\mu, \label{eq:dis22-0}\\
\epsilon_{33}(\boldk, 0^{\circ})&
=-2t_{3}(\cos k_{x}+\cos k_{y})
-4t_{4}\cos k_{x} \cos k_{y} -\mu, 
\label{eq:dis33-0}\\
\epsilon_{ab}(\boldk, 0^{\circ})&
=\ 0 \ \ \ \ \textrm{for} \ \ a\neq b \ , \label{eq:disab-0} 
\end{align} 
where $t_{1}/t_{3}=1.5$, $t_{2}/t_{3}=0.2$, and $t_{4}/t_{3}=0.4$. 
In this tight-binding model, 
we neglect both the weak hybridization 
between the $d_{xz}$ and $d_{yz}$ orbitals 
and the difference of the CEF energy between 
the $d_{xz/yz}$ and $d_{xy}$ orbitals for simplicity.~\cite{Yanase} 
In this work, 
we set $t_{3}=$ $0.45$ eV to make $W_{\textrm{tot}}$ about $4$ eV. 
Figures \ref{fig:band}{--}\ref{fig:DOS} show 
the band structure of the Ru $t_{2g}$ orbitals, the FSs, and the DOS, 
respectively. 
In the next section, 
we describe the dependence of the dispersions on $\phi$ 
for $0.5\leq x < 1.5$. 

The interacting Hamiltonian is 
\begin{align}
\hat{H}_{\textrm{int}}=& \ 
 U 
\textstyle\sum\limits_{\boldi} 
\textstyle\sum\limits_{a}
\hat{n}_{\boldi a \uparrow} \hat{n}_{\boldi a \downarrow}
+ 
U^{\prime} 
\textstyle\sum\limits_{\boldi} 
\textstyle\sum\limits_{a>b}
\hat{n}_{\boldi a} \hat{n}_{\boldi b}\notag\\
&- 
J_{\textrm{H}} 
\textstyle\sum\limits_{\boldi}  
\textstyle\sum\limits_{a>b} 
( 
2 \hat{s}^{z}_{\boldi a} \cdot 
\hat{s}^{z}_{\boldi b} 
+ 
\frac{1}{2} \hat{n}_{\boldi a} \hat{n}_{\boldi b} 
), \label{eq:Hint}
\end{align}
where $U$, $U^{\prime}$, and $J_{\textrm{H}}$ are 
the intraorbital Coulomb interaction, 
the interorbital Coulomb interaction, and 
the Hund's rule coupling; 
$\hat{\bolds}_{\boldi a}$ is defined as 
$\hat{\bolds}_{\boldi a}=
(1/2)\sum_{\sigma,\sigma^{\prime}}
\hat{c}^{\dagger}_{\boldi a \sigma} 
\boldsigma_{\sigma,\sigma^{\prime}} 
\hat{c}_{\boldi a \sigma^{\prime}}$ with  
$\boldsigma_{\sigma,\sigma^{\prime}}$ being the Pauli matrices. 
In the interacting Hamiltonian, 
we have neglected both the pair hopping $J^{\prime}$ 
and the transverse components of $J_{\textrm{H}}$ for simplicity; 
the roles of these terms are discussed in Sec. IV. 
We thus consider the following Hamiltonian in the absence of the rotation of 
RuO$_{6}$ octahedra:  
\begin{align}
&\hat{H}_{0}
+\hat{H}_{\textrm{int}}. \label{eq:red-Hint}
\end{align}

\subsection{Effects of the Ca substitution\\ on the electronic structures 
in $0.5\leq x\leq 2$}
\begin{figure}[tb]
\hspace{10pt}
\includegraphics[width=80mm]{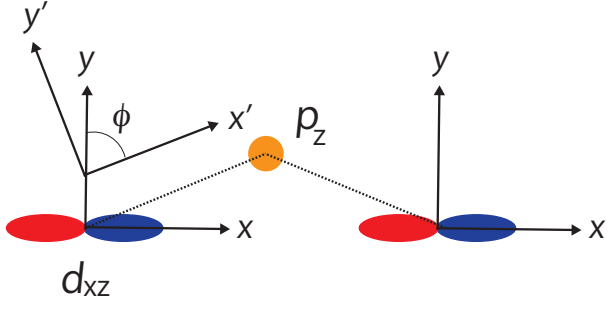}
\vspace{-8pt}
\caption{(Color online) Schematic picture of 
the in-plane $dp\pi$ hybridization of the Ru $d_{xz}$ orbital 
to the O $p_{z}$ orbital 
in the presence of the rotation of RuO$_{6}$ octahedra. 
$x$ and $y$ ($x^{\prime}$ and $y^{\prime}$) are 
the coordinates in non-rotated (rotated) frame.  
The difference of a color in the $d_{xz}$ orbital represents 
that of the sign of the wave function for the $d_{xz}$ orbital.}
\label{fig:hopping-rot}
\end{figure}
\begin{figure}[tb]
\vspace{-36pt}
\hspace{-50pt}
\includegraphics[width=102mm]{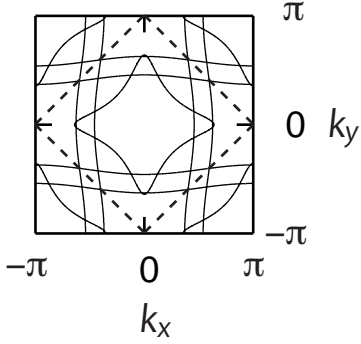}
\vspace{-30pt}
\caption{FSs for the effective model of $x=0.5$. 
The dashed line represents the folded Brillouin zone 
in the presence of the rotation of RuO$_{6}$ octahedra.}
\label{fig:FS-20-05}
\end{figure}
\begin{figure}[tb]
\vspace{6pt}
\includegraphics[width=88mm]{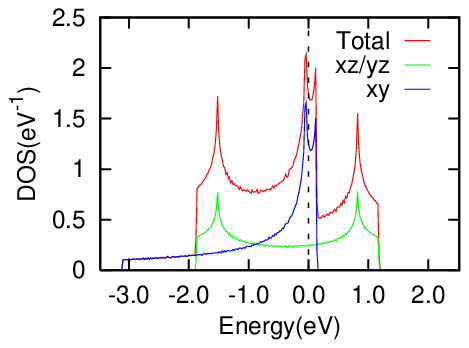}
\vspace{-16pt}
\caption{(Color online) DOS for the effective model of $x=0.5$. 
The dashed black line represents the chemical potential.}
\label{fig:DOS-20-05}
\end{figure}
\begin{figure}[tb]
\vspace{6pt}
\includegraphics[width=88mm]{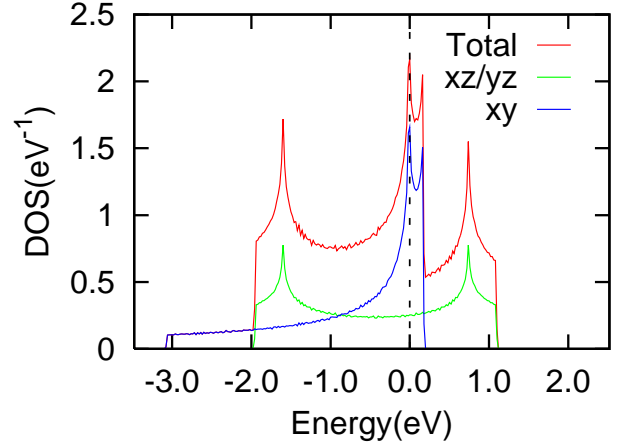}
\vspace{-16pt}
\caption{(Color online) DOS for the special model. 
The dashed black line represents the chemical potential.}
\vspace{-10pt}
\label{fig:DOS-20-vHs}
\end{figure}

As described in Sec. I, 
the Ca substitution induces the rotation of RuO$_{6}$ octahedra 
along the $c$ axis in $0.5\leq x<1.5$.  
In order to take account of effects of the Ca substitution 
on the electronic structure, 
we assume that 
the main effects of the Ca substitution are 
the changes of the $dp$ hybridizations 
due to this rotation. 
As shown in Fig. \ref{fig:rotation} (b), 
this rotation changes both the Ru-Ru lengths and 
the overlap integrals between the Ru $4d$ and O $2p$ orbitals 
keeping the Ru-O bond lengths constant. 
The former leads to negligible effects on the electronic structures 
since the angle dependence of the Ru-Ru lengths 
is $\cos \phi$, and the latter results in the change 
of the $dp$ hybridizations linear in $\phi$.  
 
We first explain the method to construct the tight-binding model 
by taking into account the effects of 
the rotation of RuO$_{6}$ octahedra on the electronic structures. 
As shown in Fig. \ref{fig:hopping-rot}, 
we can represent 
the wave functions for the Ru $4d$ and O $2p$ orbitals 
in the non-rotated frame ($x,y,z$) 
by those in the rotated frame ($x^{\prime},y^{\prime},z^{\prime}$): 
\begin{align}
\psi_{1}(\boldr)=&\ 
\psi_{d_{xz}}(\boldr)\notag\\
=&\  
\psi_{d_{x^{\prime}z^{\prime}}}(\boldr)\cos \phi 
-\psi_{d_{y^{\prime}z^{\prime}}}(\boldr)\sin \phi,\label{eq:wf-xz} \\
\psi_{2}(\boldr)=&\ 
\psi_{d_{yz}}(\boldr)\notag\\
=&\  
\psi_{d_{x^{\prime}z^{\prime}}}(\boldr)\sin \phi 
+\psi_{d_{y^{\prime}z^{\prime}}}(\boldr)\cos \phi,\label{eq:wf-yz} \\
\psi_{3}(\boldr)=&\  
\psi_{d_{xy}}(\boldr)\notag\\
=&\  
\psi_{d_{x^{\prime}y^{\prime}}}(\boldr)\cos 2\phi 
+\psi_{d_{x^{\prime 2}\textrm{-}y^{\prime 2}}}(\boldr)\sin 2\phi,\label{eq:wf-xy}\\
\psi_{4}(\boldr)=&\ 
\psi_{d_{x^{2}\textrm{-}y^{2}}}(\boldr)\notag\\
=& 
-\psi_{d_{x^{\prime}y^{\prime}}}(\boldr)\sin 2\phi 
+\psi_{d_{x^{\prime 2}\textrm{-}y^{\prime 2}}}(\boldr)\cos 2\phi,\label{eq:wf-x2-y2}
\end{align}
and
\begin{align}
\psi_{p_{x}}(\boldr)=& \ 
\psi_{p_{x^{\prime}}}(\boldr)\cos \phi 
-\psi_{p_{y^{\prime}}}(\boldr)\sin \phi,\label{eq:wf-px} \\
\psi_{p_{y}}(\boldr)=& \ 
\psi_{p_{x^{\prime}}}(\boldr)\sin \phi 
+\psi_{p_{y^{\prime}}}(\boldr)\cos \phi,\label{eq:wf-py} \\
\psi_{p_{z}}(\boldr)=& \ 
\psi_{p_{z^{\prime}}}(\boldr).\label{eq:wf-pz}
\end{align}
The $dp$ hybridizations in the rotated frame 
are equal to those at $\phi=0^{\circ}$ 
since the Ru-O bond length is unchanged;~\cite{Lattice} 
we have 
$V_{d_{x^{\prime}z^{\prime}}p_{z^{\prime}}}(\phi)=V_{1p_{z}}(0^{\circ})$ and so on. 
The finite $dp$ hybridizations at $\phi=0^{\circ}$ 
are $V_{1p_{z}}(0^{\circ})$, $V_{3p_{y}}(0^{\circ})$, and $V_{4p_{x}}(0^{\circ})$ 
along the $x$ direction 
and $V_{2p_{z}}(0^{\circ})$, $V_{3p_{x}}(0^{\circ})$, and $V_{4p_{y}}(0^{\circ})$
along the $y$ direction, respectively. 
Therefore, the $dp$ hybridizations at $\phi$ are given by 
\begin{align}
V_{1p_{z}}(\phi )&
= 
V_{1p_{z}}(0^{\circ}) \cos \phi,\\ 
V_{2p_{z}}(\phi )&
= 
V_{1p_{z}}(0^{\circ}) \sin \phi ,\\ 
V_{3p_{x}}(\phi )&
= 
V_{3p_{y}}(0^{\circ}) \sin \phi \cos 2\phi 
+ V_{4p_{x}}(0^{\circ}) \cos \phi \sin 2\phi,  \\ 
V_{3p_{y}}(\phi ) &
= 
V_{3p_{y}}(0^{\circ}) \cos \phi \cos 2\phi  
+ V_{4p_{x}}(0^{\circ}) \sin \phi \sin 2\phi , 
\end{align}
in the $x$ direction, and 
\begin{align}
V_{1p_{z}}(\phi )&
= 
-V_{2p_{z}}(0^{\circ}) \sin \phi,\\ 
V_{2p_{z}}(\phi )&
= 
V_{2p_{z}}(0^{\circ}) \cos \phi ,\\ 
V_{3p_{x}}(\phi )&
= 
V_{3p_{x}}(0^{\circ}) \cos \phi \cos 2\phi 
- V_{4p_{y}}(0^{\circ}) \sin \phi \sin 2\phi,  \\ 
V_{3p_{y}}(\phi ) &
= 
V_{3p_{x}}(0^{\circ}) \sin \phi \cos 2\phi  
+ V_{4p_{y}}(0^{\circ}) \cos \phi \sin 2\phi , 
\end{align}
in the $y$ direction, respectively. 
Using the second-order perturbation processes with respect to 
these $dp$ hybridizations, 
we obtain the following nearest-neighbor hopping integrals 
among the Ru $t_{2g}$ orbitals: 
\begin{align}
t_{11}^{A B; x}(\phi)
=&\ 
t_{1} \cos^{2} \phi ,\label{eq:hop11-x}\\ 
t_{12}^{A B; x}(\phi)
=& - \frac{1}{2} t_{1} \sin 2\phi, \label{eq:hop12-x}\\
t_{21}^{A B; x}(\phi)
=& -t_{12}^{A B; x}(\phi), \label{eq:hop21-x}\\
t_{22}^{A B; x}(\phi)
=&\ t_{2} - t_{1} \sin^{2} \phi ,\label{eq:hop22-x}\\
t_{33}^{A B; x}(\phi)
=&\ 
t_{3} \cos^{3} 2\phi 
- t_{5} \cos 2\phi \sin^{2} 2\phi\notag\\
&+ 2 t_{6} \cos 2\phi \sin^{2} 2\phi ,\label{eq:hop33-x}
\end{align}
in the $x$ direction, and 
\begin{align} 
t_{11}^{A B; y}(\phi)
=&\ t_{2} -t_{1}\sin^{2} \phi ,\label{eq:hop11-y}\\
t_{12}^{A B; y}(\phi)
=& - \frac{1}{2} t_{1} \sin 2\phi, \label{eq:hop12-y}\\
t_{21}^{A B; y}(\phi)
=& - t_{12}^{A B; y}(\phi), \label{eq:hop21-y}\\
t_{22}^{A B; y}(\phi)
=&\ t_{1} \cos^{2} \phi ,\label{eq:hop22-y}\\
t_{33}^{A B; y}(\phi)
=&\ 
t_{3} \cos^{3} 2\phi 
- t_{5} \cos 2\phi \sin^{2} 2\phi\notag\\
&+ 2 t_{6} \cos 2\phi \sin^{2} 2\phi ,\label{eq:hop33-y}
\end{align}
in the $y$ direction, respectively. 
Here, the superscripts $A$ and $B$ denote the sublattices, 
and $t_{1}$, $t_{2}$, $t_{3}$, and $t_{4}$ are defined 
in Eqs. (\ref{eq:dis11-0}){--}(\ref{eq:dis33-0}). 
In deriving Eqs. (\ref{eq:hop11-x}){--}(\ref{eq:hop33-y}), 
we have assumed that $t_{1}$ in Eqs. (\ref{eq:dis11-0}) 
and (\ref{eq:dis22-0}) arises from the second-order perturbation process 
with respect to the $dp$ hybridization 
of the Ru $d_{xz/yz}$ orbital to the O $2p_{z}$ orbital: 
$t_{1}$ is given by 
\begin{align}
t_{1}
=\dfrac{V_{1p_{z}}(0^{\circ})^{2}}{E_{1}(0^{\circ})-E_{p_{z}}(0^{\circ})}
=\dfrac{V_{2p_{z}}(0^{\circ})^{2}}{E_{2}(0^{\circ})-E_{p_{z}}(0^{\circ})}, 
\end{align}
where $E_{a}(\phi)$ and $E_{b}(\phi)$ 
are the CEF energies at $\phi$ for the Ru $t_{2g}$ orbital $a$ 
and the O $2p$ orbital $b$, respectively. 
For the case with $\phi=0^{\circ}$, 
we have $V_{3p_{x}}(0^{\circ})=V_{3p_{y}}(0^{\circ})$ 
and $E_{p_{x}}(0^{\circ})=E_{p_{y}}(0^{\circ})$. 
In the presence of the rotation of RuO$_{6}$ octahedra, 
$E_{p_{x}}(\phi)$ and $E_{p_{y}}(\phi)$ 
can be different in principle; 
however, we have assumed $E_{p_{x}}(\phi)=E_{p_{y}}(\phi)$ 
in deriving Eqs. (\ref{eq:hop11-x}){--}(\ref{eq:hop33-y}) for simplicity. 
Similarly, 
$t_{3}$ in Eq. (\ref{eq:dis33-0}) is given by
\begin{align}
t_{3}=\dfrac{V_{3p_{y}}(0^{\circ})^{2}}{E_{3}(0^{\circ})-E_{p_{y}}(0^{\circ})}
=\dfrac{V_{3p_{x}}(0^{\circ})^{2}}{E_{3}(0^{\circ})-E_{p_{x}}(0^{\circ})}. 
\end{align}
In contrast, 
$t_{2}$ in Eqs. (\ref{eq:dis11-0}) and (\ref{eq:dis22-0}) 
arises from the direct hopping process between the $d_{xz/yz}$ orbitals 
since there are no hybridizations with the O $2p$ orbitals in between; 
we have assumed that 
the rotation of RuO$_{6}$ octahedra does not affect 
$t_{2}$ in Eqs. (\ref{eq:hop22-x}) and (\ref{eq:hop11-y}). 
In Eqs. (\ref{eq:hop33-x}) and (\ref{eq:hop33-y}), 
$t_{5}$ and $t_{6}$ are the rotation-induced hopping integrals which 
are defined as 
\begin{align}
t_{5} =& 
\dfrac{V_{4p_{x}}(0^{\circ})^{2}}{E_{3}(0^{\circ})-E_{p_{x}}(\phi)}
= 
\dfrac{V_{4p_{y}}(0^{\circ})^{2}}{E_{3}(0^{\circ})-E_{p_{y}}(\phi)}, \\
t_{6} =& 
\dfrac{V_{3p_{y}}(0^{\circ})V_{4p_{x}}(0^{\circ})}
{E_{3}(0^{\circ})-E_{p_{x}}(\phi)}
=
\dfrac{V_{3p_{x}}(0^{\circ})V_{4p_{y}}(0^{\circ})}
{E_{3}(0^{\circ})-E_{p_{y}}(\phi)},
\end{align}
respectively. 
The former arises from the hybridizations of 
the $d_{x^{\prime 2}\textrm{-}y^{\prime 2}}$ orbital in Eq. (\ref{eq:wf-xy}) 
with the $p_{x^{\prime}}$ and $p_{y^{\prime}}$ orbitals
in Eqs. (\ref{eq:wf-px}) and (\ref{eq:wf-py}), and   
the latter arises from 
the hybridizations of the $d_{x^{\prime}y^{\prime}}$ 
and $d_{x^{\prime 2}\textrm{-}y^{\prime 2}}$ orbitals in Eq. (\ref{eq:wf-xy}) 
with the $p_{y^{\prime}}$ and $p_{x^{\prime}}$ orbitals 
in Eqs. (\ref{eq:wf-px}) and (\ref{eq:wf-py}). 

Using these hopping integrals Eqs. (\ref{eq:hop11-x}){--}(\ref{eq:hop33-y}), 
we obtain the noninteracting Hamiltonian 
in the presence of the rotation of RuO$_{6}$ octahedra as follows:
\begin{align}
\hat{H_{0}} = 
\sideset{}{'}\sum\limits_{\boldk}
\sideset{}{}\sum\limits_{a,b=1}^{3}
\sideset{}{}\sum\limits_{s,s^{\prime}=A,B}
\sideset{}{}\sum\limits_{\sigma}
\epsilon_{ab}^{s s^{\prime}}(\boldk, \phi)
\hat{c}^{\dagger}_{\boldk s a \sigma} 
\hat{c}_{\boldk s^{\prime} b \sigma}, 
\end{align}
where 
\begin{align}
\epsilon_{11}^{A A}(\boldk, \phi)
=& \ 
\epsilon_{22}^{A A}(\boldk, \phi)
= -\mu,\label{eq:dis11AA-phi} \\ 
\epsilon_{33}^{A A}(\boldk, \phi)
=& -4t_{4}\cos k_{x} \cos k_{y}-\mu,\label{eq:dis33AA-phi} \\
\epsilon_{11}^{A B}(\boldk, \phi)
=& \ 
-2t_{11}^{A B; x}(\phi) \cos k_{x}
-2t_{11}^{A B; y}(\phi) \cos k_{y}, 
\label{eq:dis11AB-phi}\\
\epsilon_{12}^{A B}(\boldk, \phi)
=& - 2t_{12}^{A B; x}(\phi) \cos k_{x}
- 2t_{12}^{A B; y}(\phi) \cos k_{y}, 
\label{eq:dis12AB-phi}\\
\epsilon_{21}^{A B}(\boldk, \phi)
=& - \epsilon_{12}^{A B}(\boldk, \phi), \label{eq:dis21AB-phi}\\
\epsilon_{22}^{A B}(\boldk, \phi)
=& \ 
-2t_{22}^{A B; x}(\phi) \cos k_{x}
-2t_{22}^{A B; y}(\phi) \cos k_{y}, 
\label{eq:dis22AB-phi}\\
\epsilon_{33}^{A B}(\boldk, \phi)
=& \ 
-2t_{33}^{A B; x}(\phi) \cos k_{x}
-2t_{33}^{A B; y}(\phi) \cos k_{y}, 
\label{eq:dis33AB-phi}\\
\epsilon_{ab}^{A A}(\boldk, \phi)
=& \ 
\epsilon_{ab}^{A B}(\boldk, \phi) 
= 0 
\ \ \ \ \ \ \textrm{otherwise} ,\label{eq:disab-phi}\\ 
\epsilon_{ab}^{B A}(\boldk, \phi) 
=& \ \epsilon_{ab}^{A B}(\boldk, -\phi).\label{eq:disab-BA} 
\end{align} 
The prime in the summation with respect to $\boldk$ 
represents the restriction within the folded Brillouin zone 
in the presence of the rotation of RuO$_{6}$ octahedra. 
For simplicity, 
we have neglected the effect of the rotation of RuO$_{6}$ octahedra 
on the next-nearest-neighbor hopping integrals 
for the Ru $4d$ orbitals.   

The rotation of RuO$_{6}$ octahedra also induces 
the hybridization of the $d_{xy}$ orbital to the $d_{x^{2}\textrm{-}y^{2}}$ orbital 
although it is not included in Eqs. (\ref{eq:hop11-x}){--}(\ref{eq:hop33-y}). 
Since the energy level for the $d_{x^{2}\textrm{-}y^{2}}$ orbital 
is higher than that for the $d_{xy}$ orbital, 
this hybridization leads to a downward shift of the $d_{xy}$ orbital. 
This effect is approximately taken into account as 
the difference of the CEF energy, $\Delta_{t_{2g}}$, 
between the $d_{xz/yz}$ and $d_{xy}$ orbitals: 
\begin{align}
\hat{H}_{\textrm{CEF}}= 
\dfrac{\Delta_{t_{2g}}}{3}
\textstyle\sum\limits_{\boldi}
(\hat{n}_{\boldi 1}+\hat{n}_{\boldi 2}-2\hat{n}_{\boldi 3}).
\end{align} 
In this work, we use $\Delta_{t_{2g}}$ as a parameter. 
Combining this term and the noninteracting and interacting Hamiltonians, 
we consider the following Hamiltonian 
in the presence of the rotation of RuO$_{6}$ octahedra: 
\begin{align}
&\hat{H}_{0}
+\hat{H}_{\textrm{int}}
+\hat{H}_{\textrm{CEF}}. \label{eq:H-phi}
\end{align}

In order to discuss the electronic state for $x=0.5$, 
we set $\phi=$ $15^{\circ}$ and 
($t_{5}$,$t_{6}$,$\Delta_{t_{2g}}$) $=$ ($0$,$0$,$0.42$) (eV) 
so as to reproduce the FSs obtained in the ARPES measurement.~\cite{ARPES05} 
Figures \ref{fig:FS-20-05} and \ref{fig:DOS-20-05} 
represent the FSs and the DOS, respectively. 
In Fig. \ref{fig:FS-20-05}, 
the absence of the FS around $\boldk=(\pi,0)$ 
corresponds to the change of the FS for the $d_{xy}$ orbital 
from an electron pocket to a hole pocket. 
There are three main changes of the electronic structures 
due to the rotation of RuO$_{6}$ octahedra: 
the reduction of the bandwidth for the $d_{xy}$ orbital, 
the increase of the pDOS for the $d_{xy}$ orbital near the Fermi level, 
and the change of the FS for the $d_{xy}$ orbital 
from an electron pocket to a hole pocket. 
The bandwidth and pDOS for the $d_{xz/yz}$ orbital are little affected. 
These results are qualitatively consistent with those obtained 
in the density-functional calculation 
within the LDA~\cite{Terakura} or 
local-spin-density approximation.~\cite{Oguchi-LS} 
We thus think that the present model is reasonable 
for the study of the electronic states in Ca$_{2-x}$Sr$_{x}$RuO$_{4}$ 
although the experimental value of $\phi$ is about $12.8^{\circ}$ 
at $x=0.5$.~\cite{Lattice}

In addition, 
we study a special model in which 
the vHs for the $d_{xy}$ orbital is located on the Fermi level 
in order to clarify the role of the vHs in enhancing the effective mass. 
We use the dispersions in Eqs. (\ref{eq:dis11AA-phi}){--}(\ref{eq:disab-BA}) 
setting $\phi=$ $15^{\circ}$ and 
($t_{5}$,$t_{6}$,$\Delta_{t_{2g}}$) $=$ ($0$,$0$,$0.3$) (eV); 
the difference between this case and the case of $x=0.5$ 
is that in the values of $\Delta_{t_{2g}}$. 
Figure \ref{fig:DOS-20-vHs} shows the DOS in this case. 

\subsection{Gutzwiller approximation}
In this section, 
we explain the GA for a PM state consisting of 
the degenerate $d_{xz}$ and $d_{yz}$ orbitals and the $d_{xy}$ orbital. 
Let us define the Gutzwiller-type variational wave function as  
\begin{align}
|\Psi_{\textrm{G}} \rangle 
=
\prod_{\boldj}
\prod_{l=0}^{63} 
\bigl[1-(1-g_{l}) \hat{P}_{\boldj;l}\bigr] 
|\Psi_{0} \rangle , 
\end{align}
where $|\Psi_{0} \rangle$ is the ground state for $\hat{H}_{0}$,  
$\hat{P}_{\boldj;l}$ are the projection operators, 
and $g_{l}$ are the corresponding variational parameters 
for the $l$th configuration. 
There are $4^{3}=64$ configurations for each site $\boldj$ 
since we have three orbitals and two kinds of spin degrees of freedom. 
The variational energy is given by  
\begin{align}
E_{\textrm{gs}}=
\min_{\{g_{l}\}} 
\dfrac{\langle \Psi_{\textrm{G}} | \hat{H} |\Psi_{\textrm{G}} \rangle}
{\langle \Psi_{\textrm{G}}  |\Psi_{\textrm{G}} \rangle}, \label{eq:Egs}
\end{align}
where $\min_{\{g_{l}\}}$ denotes the optimization with respect to ${\{g_{l}\}}$, 
and $\hat{H}$ is the total Hamiltonian given by Eq. (\ref{eq:H-phi}). 

To calculate the expectation values, 
we use the GA~\cite{GA,GA-Ogawa,GA-Fazekas,GA-multi,GA-tJ} in which 
the dependence of the Slater determinants on the configurations 
is neglected. 
For example, the square of the Slater determinant, 
$[\textrm{det}\ U_{1\uparrow}(\{\boldr_{1\uparrow}^{N_{1\uparrow}}\})]^{2}$, 
for the electrons in $a=1$ with $\sigma=\ \uparrow$ 
can be approximated as 
\begin{align}
\bigl[\textrm{det}\ U_{1\uparrow}(\{\boldr_{1\uparrow}^{N_{1\uparrow}}\})\bigr]^{2}
\stackrel{\textrm{GA}}{\to} 
(n_{1\uparrow}^{0})^{N_{1\uparrow}}
(1- n_{1\uparrow}^{0})^{L-N_{1\uparrow}},  
\end{align}
where $\{\boldr_{1\uparrow}^{N_{1\uparrow}}\}$ is a set of sites occupied 
by these electrons, 
$L$ is the number of sites, 
$N_{1\uparrow}$ is the number of these electrons, 
and $n_{1\uparrow}^{0}$ is defined as 
an expectation value without the projection operators: 
\begin{align} 
n_{1\uparrow}^{0}
=
\dfrac{1}{L} 
\textstyle\sum\limits_{\boldj}
\langle \Psi_{0}| 
\hat{n}_{\boldj 1\uparrow} 
|\Psi_{0} \rangle 
= 
\dfrac{1}{L} 
\textstyle\sum\limits_{\boldj}
\langle \hat{n}_{\boldj 1 \uparrow} \rangle_{0}.
\end{align}

Using the GA, 
we can rewrite the denominator in Eq. (\ref{eq:Egs}) as 
\begin{align}
\langle \Psi_{\textrm{G}}|\Psi_{\textrm{G}} \rangle 
&\ =
\textstyle\sum\limits_{\{N_{a\sigma}\}}
\textstyle\sum\limits_{\{\boldr_{a\sigma}^{N_{a\sigma}}\}} 
\bigl[\textstyle\prod\limits_{l}g_{l}^{2\Gamma_{l}(\{\boldr_{a\sigma}^{N_{a\sigma}}\})}\bigr]
\notag\\
&\ \ \ \ \times 
\textstyle\prod\limits_{a=1}^{3}
\textstyle\prod\limits_{\sigma}
\bigl[\textrm{det}\ U_{a\sigma}(\{\boldr_{a\sigma}^{N_{a\sigma}}\})\bigr]^{2} 
\notag\\
&\stackrel{\textrm{GA}}{\to} 
\textstyle\sum\limits_{\{N_{a\sigma}\}}
\textstyle\sum\limits_{\{\Gamma_{l}\}}
\bigl(\textstyle\prod\limits_{l}g_{l}^{2\Gamma_{l}}\bigr) 
\dfrac{L!}{(\textstyle\prod\limits_{l}\Gamma_{l}!)}
P(L;\{N_{a\sigma}\}) , \label{eq:denom}
\end{align}
with
\begin{align}
P(L;\{N_{a\sigma}\})&= 
\prod_{a=1}^{3}
\prod_{\sigma}
(n_{a\sigma}^{0})^{N_{a\sigma}}
(1-n_{a\sigma}^{0})^{L-N_{a\sigma}}. 
\end{align}
Here, $\sum_{\{N_{a\sigma}\}}$ denotes the summation with respect to all 
the possible values $N_{a\sigma}$ ($a=1,2,3$, $\sigma=\ \uparrow, \downarrow$), 
$\Gamma_{l}$ is the number of sites with the $l$th configuration, 
and the summation $\sum_{\{\Gamma_{l}\}}$ is taken over 
all the possible values $\Gamma_{l}$ ($l=0-63$) under the constraints
\begin{align}
\textstyle\sum\limits_{l=0}^{63} \Gamma_{l}=&\ L,\ \label{eq:constraint-0}\\  
\textstyle\sum\limits_{l=0}^{63}
\textstyle\sum\limits_{a=1,2} \Gamma_{l}n_{la\sigma} =& 
\textstyle\sum\limits_{a=1,2}N_{a\sigma},\label{eq:constraint-1}\\  
\textstyle\sum\limits_{l=0}^{63} \Gamma_{l}n_{l3\sigma} =&\ N_{3\sigma}, \label{eq:constraint-2}
\end{align}  
where $n_{la\sigma}$ is the number of electrons in orbital $a$ 
with spin $\sigma$ for the $l$th configuration; 
as we shall show, $n_{la\sigma}$ plays important roles 
in calculating the expectation values within the GA. 

Since the occupation number for each orbital is same 
in $A$ and $B$ sublattices, 
we do not need to take care of the sublattices in the GA. 
In other words,  
the formulation for a PM state consisting of the $t_{2g}$ orbitals 
without the sublattice indices 
is sufficient to calculate the expectation values 
for a PM state consisting of the $t_{2g}$ orbitals 
with the sublattice indices. 

$(N_{1\sigma}+N_{2\sigma})$ and $N_{3\sigma}$ are conserved quantum numbers 
although $N_{1\sigma}$ and $N_{2\sigma}$ are not conserved quantum numbers 
due to the finite hopping integral between the $d_{xz}$ and $d_{yz}$ orbitals. 
In general, the values of $(N_{1\sigma}+N_{2\sigma})$ and $N_{3\sigma}$ may change 
in the presence of interactions 
for a system with orbital degrees of freedom. 
In the following, however, 
we assume that these values are equal to 
those in the absence of the interactions; 
e.g., for the effective model of $x=2$, 
$(N_{1\sigma}+N_{2\sigma})/L=(n_{1}^{0}+n_{2}^{0})=2.66$,
$N_{3\sigma}/L=n_{3}^{0}=1.33$.
In Sec. IV, 
we address the effect of the variation of the occupation numbers 
due to electron correlation.  

In the thermodynamic limit, 
we can approximate the summation 
with respect to $\Gamma_{l}$ in Eqs. (\ref{eq:denom}) to their largest terms:  
\begin{align}
&\langle \Psi_{\textrm{G}}|\Psi_{\textrm{G}} \rangle  
\stackrel{\textrm{GA}}{\to}   
\Bigl(\textstyle\prod\limits_{l} 
g_{l}^{2\bar{\Gamma}_{l}}\Bigr) 
\dfrac{L!}{(\textstyle\prod\limits_{l}\bar{\Gamma}_{l}!)}
P(L;\{N_{a\sigma}\}) ,\label{eq:denom-GA}
\end{align}
where $\{\bar{\Gamma}_{l}\}$ is the set of $\{\Gamma_{l}\}$ 
that gives the largest term. 
$\bar{\Gamma}_{l}$ is given by
\begin{align}
\bar{\Gamma}_{l}=&\ L g_{l}^{2}
\exp\bigl(1+\lambda_{0}
+\textstyle\sum\limits_{a=1}^{3}
\textstyle\sum\limits_{\sigma}\lambda_{a\sigma}n_{la\sigma}
\bigr),
\label{eq:X-g}
\end{align}
where ($\lambda_{0}$, $\{\lambda_{a\sigma}\}$) 
are the Lagrange multipliers~\cite{GA-tJ} determined 
so as to satisfy the constraints 
Eqs. (\ref{eq:constraint-0}){--}(\ref{eq:constraint-2}) 
for $\{\bar{\Gamma}_{l}\}$. 
For simplicity, we assume that 
the following relations hold in the thermodynamic limit: 
\begin{align}
\dfrac{N_{a\uparrow}}{L}=&\ \dfrac{N_{a\downarrow}}{L},\label{eq:N-sym2}\\
\dfrac{N_{1\sigma}}{L}=&\ \dfrac{N_{2\sigma}}{L}.\label{eq:N-sym1}
\end{align}
Correspondingly, 
the Lagrange multipliers satisfy 
\begin{align}
\lambda_{a\uparrow}=&\ \lambda_{a\downarrow},\label{eq:Lag-sym1}\\
\lambda_{2\sigma}=&\ \lambda_{1\sigma}.\label{eq:Lag-sym2} 
\end{align}

Similarly, 
the expectation values of 
$\hat{n}_{\boldi a \sigma}\hat{n}_{\boldi b \sigma^{\prime}}$ and 
$\hat{c}_{\boldi a \sigma}^{\dagger}\hat{c}_{\boldj b \sigma}$ 
can be rewritten within the GA as 
\begin{align}
&
\langle \Psi_{\textrm{G}}| 
\hat{n}_{\boldi a \sigma}\hat{n}_{\boldi b \sigma^{\prime}} 
|\Psi_{\textrm{G}} \rangle\notag\\
\stackrel{\textrm{GA}}{\to}&\  
\sideset{}{'}\sum\limits_{l_{\boldi}}
\sum\limits_{\{\Gamma_{l}^{\prime}\}}
g_{l_{\boldi}}^{2} 
\langle \hat{P}_{\boldi;l_{\boldi}} \hat{n}_{\boldi a \sigma}
\hat{n}_{\boldi b \sigma^{\prime}} \hat{P}_{\boldi;l_{\boldi}} \rangle_{0}
\Bigl(\prod\limits_{l}g_{l}^{2\Gamma_{l}^{\prime}}\Bigr)\notag\\[-2pt]
&\times 
\dfrac{(L-1)!}{(\textstyle\prod\limits_{l}\Gamma_{l}^{\prime}!)}
P(L-1;\{N_{a\sigma}^{\prime}\}), \label{eq:expect-int-before}
\end{align}
\vspace{-4pt}
and 
\begin{align}
&
\langle \Psi_{\textrm{G}}| 
\hat{c}_{\boldi a \sigma}^{\dagger}\hat{c}_{\boldj b \sigma} 
|\Psi_{\textrm{G}} \rangle \notag\\
\stackrel{\textrm{GA}}{\to}&\ 
\sideset{}{''}\sum\limits_{l_{\boldi},l_{\boldj},l_{\boldi}^{\prime},l_{\boldj}^{\prime}}
\sum\limits_{\{\Gamma_{l}^{\prime\prime}\}}
g_{l_{\boldi}}g_{l_{\boldj}}g_{l_{\boldi}^{\prime}}g_{l_{\boldj}^{\prime}} 
\langle \hat{P}_{\boldi;l_{\boldi}}\hat{P}_{\boldj;l_{\boldj}} 
\hat{c}_{\boldi a \sigma}^{\dagger}
\hat{c}_{\boldj b \sigma} 
\hat{P}_{\boldi;l_{\boldi}^{\prime}}\hat{P}_{\boldj;l_{\boldj}^{\prime}}
\rangle_{0}\notag\\[-2pt]
&\times 
\Bigl(\prod\limits_{l}g_{l}^{2\Gamma_{l}^{\prime\prime}}\Bigr)
\dfrac{(L-2)!}{(\textstyle\prod\limits_{l}\Gamma_{l}^{\prime\prime}!)}
P(L-2;\{N_{a\sigma}^{\prime\prime}\}), \label{eq:expect-kinetic-before}
\end{align}
with the constraints
\begin{align}
\textstyle\sum\limits_{l=0}^{63} \Gamma_{l}^{\prime}
=&\ L-1,\label{eq:constraint-prime1-0}\\  
\textstyle\sum\limits_{l=0}^{63}
\textstyle\sum\limits_{a=1,2}  
\Gamma_{l}^{\prime}n_{la\sigma}
=&\ \textstyle\sum\limits_{a=1,2} N_{a\sigma}^{\prime}\notag\\
=&\ \textstyle\sum\limits_{a=1,2} (N_{a\sigma}-n_{l_{\boldi}a\sigma})
,\label{eq:constraint-prime1-1}\\
\textstyle\sum\limits_{l=0}^{63} 
\Gamma_{l}^{\prime}n_{l3\sigma}
=&\ N_{3\sigma}^{\prime}\notag\\
=&\ N_{3\sigma}-n_{l_{\boldi}3\sigma}
,\label{eq:constraint-prime1-2}
\end{align}
and
\begin{align}
\textstyle\sum\limits_{l=0}^{63} \Gamma_{l}^{\prime\prime}
=&\ L-2,\label{eq:constraint-prime2-0}\\  
\textstyle\sum\limits_{l=0}^{63} 
 \textstyle\sum\limits_{a=1,2}
\Gamma_{l}^{\prime\prime}n_{la\sigma} 
=& \textstyle\sum\limits_{a=1,2}N_{a\sigma}^{\prime\prime}\notag\\
=& \textstyle\sum\limits_{a=1,2}
(N_{a\sigma}-n_{l_{\boldi}a\sigma}-n_{l_{\boldj}a\sigma})\notag\\
=& \textstyle\sum\limits_{a=1,2}
(N_{a\sigma}-n_{l_{\boldi}^{\prime}a\sigma}-n_{l_{\boldj}^{\prime}a\sigma}),
\label{eq:constraint-prime2-1}\\
\textstyle\sum\limits_{l=0}^{63} 
\Gamma_{l}^{\prime\prime}n_{l3\sigma} 
=&\ N_{3\sigma}^{\prime\prime}\notag\\
=&\ N_{3\sigma}-n_{l_{\boldi}3\sigma}-n_{l_{\boldj}3\sigma}\notag\\
=&\ N_{3\sigma}-n_{l_{\boldi}^{\prime}3\sigma}-n_{l_{\boldj}^{\prime}3\sigma}.
\label{eq:constraint-prime2-2} 
\end{align}  
Here, the summation with respect to $l_{\boldi}$ 
in Eq. (\ref{eq:expect-int-before}) 
is taken for the configurations of site $\boldi$ in which 
there are at least an electron in orbital $a$ with spin $\sigma$ 
and an electron in orbital $b$ with spin $\sigma^{\prime}$. 
In Eq. (\ref{eq:expect-kinetic-before}), 
$l_{\boldi}$ and $l_{\boldj}$ are the configurations 
of sites $\boldi$ and $\boldj$ 
after the hopping process $\hat{c}_{\boldi a \sigma}^{\dagger}\hat{c}_{\boldj b \sigma}$ 
in which there are no electrons in orbital $b$ with spin $\sigma$ 
at site $\boldj$ and there is at least an electron 
in orbital $a$ with spin $\sigma$ at site $\boldi$; 
$l_{\boldi}^{\prime}$ and $l_{\boldj}^{\prime}$ are similar configurations 
of sites $\boldi$ and $\boldj$ before the hopping process. 
The prime and double prime in the summations of 
Eqs. (\ref{eq:expect-int-before}) and (\ref{eq:expect-kinetic-before}) 
represent these restrictions. 

In the thermodynamic limit, 
we approximate 
the summation with respect to $\Gamma_{l}^{\prime}$ in 
Eq. (\ref{eq:expect-int-before}) to their largest terms 
for each configuration of $l_{\boldi}$ 
and the summation with respect to $\Gamma_{l}^{\prime\prime}$ in 
Eq. (\ref{eq:expect-kinetic-before}) to their largest terms 
for the configurations of $l_{\boldi}$, $l_{\boldj}$, $l_{\boldi}^{\prime}$, 
and $l_{\boldj}^{\prime}$. 
Thus, these expectation values become 
\begin{align}
&
\langle \Psi_{\textrm{G}}| 
\hat{n}_{\boldi a \sigma}\hat{n}_{\boldi b \sigma^{\prime}} 
|\Psi_{\textrm{G}} \rangle\notag\\
\stackrel{\textrm{GA}}{\to}&\ 
\sideset{}{'}\sum\limits_{l_{\boldi}}
g_{l_{\boldi}}^{2} 
\langle \hat{P}_{\boldi;l_{\boldi}} \hat{n}_{\boldi a \sigma}
\hat{n}_{\boldi b \sigma^{\prime}} \hat{P}_{\boldi;l_{\boldi}} \rangle_{0}
\Bigl(\prod\limits_{l}g_{l}^{2\bar{\Gamma}_{l}^{\prime}}\Bigr)\notag\\[-2pt]
&\times 
\dfrac{(L-1)!}{(\textstyle\prod\limits_{l}\bar{\Gamma}_{l}^{\prime}!)}
P(L-1;\{N_{a\sigma}^{\prime}\}), \label{eq:expect-int}
\end{align}
\vspace{-4pt}
and
\begin{align}
&
\langle \Psi_{\textrm{G}}| 
\hat{c}_{\boldi a \sigma}^{\dagger}\hat{c}_{\boldj b \sigma} 
|\Psi_{\textrm{G}} \rangle \notag\\
\stackrel{\textrm{GA}}{\to}&\ 
\sideset{}{''}\sum\limits_{l_{\boldi},l_{\boldj},l_{\boldi}^{\prime},l_{\boldj}^{\prime}}
g_{l_{\boldi}}g_{l_{\boldj}}g_{l_{\boldi}^{\prime}}g_{l_{\boldj}^{\prime}} 
\langle \hat{P}_{\boldi;l_{\boldi}}\hat{P}_{\boldj;l_{\boldj}} 
\hat{c}_{\boldi a \sigma}^{\dagger}
\hat{c}_{\boldj b \sigma} 
\hat{P}_{\boldi;l_{\boldi}^{\prime}}\hat{P}_{\boldj;l_{\boldj}^{\prime}}
\rangle_{0}\notag\\[-2pt]
&\times 
\Bigl(\prod\limits_{l}g_{l}^{2\bar{\Gamma}_{l}^{\prime\prime}}\Bigr)
\dfrac{(L-2)!}{(\textstyle\prod\limits_{l}\bar{\Gamma}_{l}^{\prime\prime}!)}
P(L-2;\{N_{a\sigma}^{\prime\prime}\}), \label{eq:expect-kinetic}
\end{align}
where 
$\bar{\Gamma}_{l}^{\prime}$ and $\bar{\Gamma}_{l}^{\prime\prime}$ are given by 
\begin{align}
\bar{\Gamma}_{l}^{\prime}=&\ L g_{l}^{2}
\exp\bigl(1+\lambda_{0}^{\prime}
+\textstyle\sum\limits_{a=1}^{3}
\textstyle\sum\limits_{\sigma}\lambda_{a\sigma}^{\prime}n_{la\sigma}
\bigr),
\label{eq:X-g-int}\\
\bar{\Gamma}_{l}^{\prime\prime}=&\ L g_{l}^{2}
\exp\bigl(1+\lambda_{0}^{\prime\prime}
+\textstyle\sum\limits_{a=1}^{3}
\textstyle\sum\limits_{\sigma}
\lambda_{a\sigma}^{\prime\prime}n_{la\sigma}\bigr),
\label{eq:X-g-KE}
\end{align}
respectively. 
The Lagrange multipliers, 
($\lambda_{0}^{\prime}$, $\{\lambda_{a\sigma}^{\prime}\}$) 
and ($\lambda_{0}^{\prime\prime}$, $\{\lambda_{a\sigma}^{\prime\prime}\}$), 
are determined so as to satisfy the constraints 
Eqs. (\ref{eq:constraint-prime1-0}){--}(\ref{eq:constraint-prime1-2}) 
for $\{\bar{\Gamma}_{l}^{\prime}\}$ 
and Eqs. (\ref{eq:constraint-prime2-0}){--}(\ref{eq:constraint-prime2-2}) 
for $\{\bar{\Gamma}_{l}^{\prime\prime}\}$, 
respectively. 
For simplicity, we also assume that 
the following relations hold in the thermodynamic limit: 
\begin{align}
\dfrac{N_{a\uparrow}^{\prime}}{L}=&\ \dfrac{N_{a\downarrow}^{\prime}}{L},\label{eq:Np-sym2}\\
\dfrac{N_{1\sigma}^{\prime}}{L}=&\ \dfrac{N_{2\sigma}^{\prime}}{L},\label{eq:Np-sym1}
\end{align}
and
\begin{align}
\dfrac{N_{a\uparrow}^{\prime\prime}}{L}=&\ \dfrac{N_{a\downarrow}^{\prime\prime}}{L},\label{eq:Npp-sym2}\\
\dfrac{N_{1\sigma}^{\prime\prime}}{L}=&\ \dfrac{N_{2\sigma}^{\prime\prime}}{L}.\label{eq:Npp-sym1}
\end{align}
Correspondingly, the Lagrange multipliers satisfy 
\begin{align}
\lambda_{a\uparrow}^{\prime}=&\ \lambda_{a\downarrow}^{\prime},\label{eq:Lagp-sym1}\\
\lambda_{2\sigma}^{\prime}=&\ \lambda_{1\sigma}^{\prime},\label{eq:Lagp-sym2}
\end{align}
and
\begin{align}
\lambda_{a\uparrow}^{\prime\prime}=&\ \lambda_{a\downarrow}^{\prime\prime},\label{eq:Lagpp-sym1}\\
\lambda_{2\sigma}^{\prime\prime}=&\ \lambda_{1\sigma}^{\prime\prime}.\label{eq:Lagpp-sym2}  
\end{align}

Using Eqs. (\ref{eq:denom-GA}), 
(\ref{eq:expect-int}), and (\ref{eq:expect-kinetic}), 
we obtain the normalized expectation values 
of $\hat{n}_{\boldi a \sigma}\hat{n}_{\boldi b \sigma^{\prime}}$ and 
$\hat{c}_{\boldi a \sigma}^{\dagger}\hat{c}_{\boldj b \sigma}$ 
within the GA in the thermodynamic limit:
\begin{align}
&
\dfrac{\langle \Psi_{\textrm{G}}| 
\hat{n}_{\boldi a \sigma}\hat{n}_{\boldi b \sigma^{\prime}} 
|\Psi_{\textrm{G}} \rangle}
{\langle \Psi_{\textrm{G}}  |\Psi_{\textrm{G}} \rangle}\notag\\[-2pt]
\stackrel{\textrm{GA}}{\to}&\ 
\frac{1}{L}
\sideset{}{'}\sum\limits_{l_{\boldi}}
g_{l_{\boldi}}^{2} 
\langle \hat{P}_{\boldi;l_{\boldi}} \hat{n}_{\boldi a \sigma}
\hat{n}_{\boldi b \sigma^{\prime}} \hat{P}_{\boldi;l_{\boldi}} \rangle_{0} \notag\\[-5pt]
&\times 
\Bigl(\textstyle\prod\limits_{l}g_{l}^{2\Delta\bar{\Gamma}_{l}^{\prime}}\Bigr)
\dfrac{(\textstyle\prod\limits_{l}\bar{\Gamma}_{l}!)}
{(\textstyle\prod\limits_{l}\bar{\Gamma}_{l}^{\prime}!)}
\dfrac{P(L-1;\{N_{a\sigma}^{\prime}\})}{P(L;\{N_{a\sigma}\})},
\label{eq:normalized-1}
\end{align}
and 
\begin{align}
&
\dfrac{\langle \Psi_{\textrm{G}}| 
\hat{c}_{\boldi a \sigma}^{\dagger}\hat{c}_{\boldj b \sigma}  
|\Psi_{\textrm{G}} \rangle}
{\langle \Psi_{\textrm{G}}  |\Psi_{\textrm{G}} \rangle}\notag\\[-2pt]
\stackrel{\textrm{GA}}{\to}&\ 
\frac{1}{L^{2}}
\sideset{}{''}\sum\limits_{l_{\boldi},l_{\boldj},l_{\boldi}^{\prime},l_{\boldj}^{\prime}}
g_{l_{\boldi}}g_{l_{\boldj}}g_{l_{\boldi}^{\prime}}g_{l_{\boldj}^{\prime}} 
\langle \hat{P}_{\boldi;l_{\boldi}}\hat{P}_{\boldj;l_{\boldj}} 
\hat{c}_{\boldi a \sigma}^{\dagger}
\hat{c}_{\boldj b \sigma} 
\hat{P}_{\boldi;l_{\boldi}^{\prime}}\hat{P}_{\boldj;l_{\boldj}^{\prime}}
\rangle_{0}\notag\\[-5pt]
&\times 
\Bigl(\textstyle\prod\limits_{l}g_{l}^{2\Delta\bar{\Gamma}_{l}^{\prime\prime}}\Bigr) 
\dfrac{(\textstyle\prod\limits_{l}\bar{\Gamma}_{l}!)}
{(\textstyle\prod\limits_{l}\bar{\Gamma}_{l}^{\prime\prime}!)}
\dfrac{P(L-2;\{N_{a\sigma}^{\prime\prime}\})}{P(L;\{N_{a\sigma}\})},\label{eq:normalized-2} 
\end{align} 
where $\Delta\bar{\Gamma}_{l}^{\prime}$ 
and $\Delta\bar{\Gamma}_{l}^{\prime\prime}$ are defined as 
$\Delta\bar{\Gamma}_{l}^{\prime}=\bar{\Gamma}_{l}^{\prime}-\bar{\Gamma}_{l}$ 
and $\Delta\bar{\Gamma}_{l}^{\prime\prime}
=\bar{\Gamma}_{l}^{\prime\prime}-\bar{\Gamma}_{l}$, respectively. 
From Eqs. (\ref{eq:constraint-0}){--}(\ref{eq:constraint-2}) 
for $\{\bar{\Gamma}_{l}\}$, 
Eqs. (\ref{eq:constraint-prime1-0}){--}(\ref{eq:constraint-prime1-2}) 
for $\{\bar{\Gamma}_{l}^{\prime}\}$, and 
Eqs. (\ref{eq:constraint-prime2-0}){--}(\ref{eq:constraint-prime2-2}) 
for $\{\bar{\Gamma}_{l}^{\prime\prime}\}$, 
we have the following constraints for 
$\Delta\bar{\Gamma}_{l}^{\prime}$ and $\Delta\bar{\Gamma}_{l}^{\prime\prime}$: 
\begin{align}
\textstyle\sum\limits_{l=0}^{63} \Delta \bar{\Gamma}_{l}^{\prime}
=&\ -1,\label{eq:constraint-devi-int0}\\  
\textstyle\sum\limits_{l=0}^{63}\textstyle\sum\limits_{a=1,2} 
\Delta \bar{\Gamma}_{l}^{\prime}n_{la\sigma}
=&\ -\textstyle\sum\limits_{a=1,2}n_{l_{\boldi}a\sigma},\label{eq:constraint-devi-int1}\\
\textstyle\sum\limits_{l=0}^{63} \Delta \bar{\Gamma}_{l}^{\prime}n_{l3\sigma}
=&\ -n_{l_{\boldi}3\sigma},\label{eq:constraint-devi-int2}
\end{align}
and
\begin{align}
\textstyle\sum\limits_{l=0}^{63} \Delta \bar{\Gamma}_{l}^{\prime\prime}
=&\ -2,\label{eq:constraint-devi-kinetic0}\\
\textstyle\sum\limits_{l=0}^{63}\textstyle\sum\limits_{a=1,2}
 \Delta \bar{\Gamma}_{l}^{\prime\prime}n_{la\sigma}
=&\ -\textstyle\sum\limits_{a=1,2}(n_{l_{\boldi}a\sigma}+n_{l_{\boldj}a\sigma})\notag\\
=&\ -\textstyle\sum\limits_{a=1,2}(n_{l_{\boldi}^{\prime}a\sigma}+n_{l_{\boldj}^{\prime}a\sigma}),
\label{eq:constraint-devi-kinetic1}\\
\textstyle\sum\limits_{l=0}^{63} \Delta \bar{\Gamma}_{l}^{\prime\prime}n_{l3\sigma}
=&\ -(n_{l_{\boldi}3\sigma}+n_{l_{\boldj}3\sigma})\notag\\
=&\ -(n_{l_{\boldi}^{\prime}3\sigma}+n_{l_{\boldj}^{\prime}3\sigma}).\label{eq:constraint-devi-kinetic2}
\end{align}
Calculating 
$\langle \hat{P}_{\boldi;l_{\boldi}} \hat{n}_{\boldi a \sigma}
\hat{n}_{\boldi b \sigma^{\prime}} \hat{P}_{\boldi;l_{\boldi}} \rangle_{0}$ and 
$\frac{P(L-1;\{N_{a\sigma}^{\prime}\})}{P(L;\{N_{a\sigma}\})}$ 
in Eq. (\ref{eq:normalized-1}) and 
$\langle \hat{P}_{\boldi;l_{\boldi}}\hat{P}_{\boldj;l_{\boldj}} 
\hat{c}_{\boldi a \sigma}^{\dagger}
\hat{c}_{\boldj b \sigma} 
\hat{P}_{\boldi;l_{\boldi}^{\prime}}\hat{P}_{\boldj;l_{\boldj}^{\prime}}
\rangle_{0}$ and 
$\frac{P(L-2;\{N_{a\sigma}^{\prime\prime}\})}{P(L;\{N_{a\sigma}\})}$ 
in Eq. (\ref{eq:normalized-2}) 
explicitly for each possible configurations, 
we find
\begin{align}
&\dfrac{\langle \Psi_{\textrm{G}}| 
\hat{n}_{\boldi a \sigma}\hat{n}_{\boldi b \sigma^{\prime}} 
|\Psi_{\textrm{G}} \rangle}
{\langle \Psi_{\textrm{G}}  |\Psi_{\textrm{G}} \rangle}
\stackrel{\textrm{GA}}{\to}\ 
\frac{1}{L}
\sideset{}{'}\sum\limits_{l_{\boldi}}
g_{l_{\boldi}}^{2} 
\Bigl(\textstyle\prod\limits_{l}g_{l}^{2\Delta\bar{\Gamma}_{l}^{\prime}}\Bigr) 
\dfrac{\bigl(\textstyle\prod\limits_{l}\bar{\Gamma}_{l}!\bigr)}
{\bigl(\textstyle\prod\limits_{l}\bar{\Gamma}_{l}^{\prime}!\bigr)},
\label{eq:normalized-1-ver2}
\end{align}
and
\begin{widetext}
\begin{align}
\dfrac{\langle \Psi_{\textrm{G}}| 
\hat{c}_{\boldi a \sigma}^{\dagger}\hat{c}_{\boldj b \sigma}  
|\Psi_{\textrm{G}} \rangle}
{\langle \Psi_{\textrm{G}}  |\Psi_{\textrm{G}} \rangle}
\stackrel{\textrm{GA}}{\to}\ 
\frac{1}{L^{2}}
\sideset{}{''}\sum\limits_{l_{\boldi},l_{\boldj},l_{\boldi}^{\prime},l_{\boldj}^{\prime}}
g_{l_{\boldi}}g_{l_{\boldj}}g_{l_{\boldi}^{\prime}}g_{l_{\boldj}^{\prime}} 
\langle  
\hat{c}_{\boldi a \sigma}^{\dagger}
\hat{c}_{\boldj b \sigma} 
\rangle_{0}
\Bigl(\textstyle\prod\limits_{l}g_{l}^{2\Delta\bar{\Gamma}_{l}^{\prime\prime}}\Bigr) 
\dfrac{\bigl(\textstyle\prod\limits_{l}\bar{\Gamma}_{l}!\bigr)}
{\bigl(\textstyle\prod\limits_{l}\bar{\Gamma}_{l}^{\prime\prime}!\bigr)}
\dfrac{1}
{(1-n_{a}^{0})n_{b}^{0}}.\label{eq:normalized-2-ver2} 
\end{align} 
In addition, we have the following relation in the thermodynamic limit: 
\begin{align}
\dfrac{(\bar{\Gamma}_{l})!}{(\bar{\Gamma}_{l}+\Delta \bar{\Gamma}_{l})!}
\sim(\bar{\Gamma}_{l})^{-\Delta \bar{\Gamma}_{l}}.\label{eq:formula1}
\end{align}
Using this relation, we can rewrite Eqs. (\ref{eq:normalized-1-ver2}) and 
(\ref{eq:normalized-2-ver2}) as
\begin{align}
\dfrac{\langle \Psi_{\textrm{G}}| 
\hat{n}_{\boldi a \sigma}\hat{n}_{\boldi b \sigma^{\prime}} 
|\Psi_{\textrm{G}} \rangle}
{\langle \Psi_{\textrm{G}}  |\Psi_{\textrm{G}} \rangle}
&\stackrel{\textrm{GA}}{\to}\ 
\frac{1}{L}
\sideset{}{'}\sum\limits_{l_{\boldi}}
g_{l_{\boldi}}^{2} 
\textstyle\prod\limits_{l}
\biggl(\dfrac{\bar{\Gamma}_{l}}{g_{l}^{2}}\biggr)^{-\Delta\bar{\Gamma}_{l}^{\prime}},\\
\dfrac{\langle \Psi_{\textrm{G}}| 
\hat{c}_{\boldi a \sigma}^{\dagger}\hat{c}_{\boldj b \sigma}  
|\Psi_{\textrm{G}} \rangle}
{\langle \Psi_{\textrm{G}}  |\Psi_{\textrm{G}} \rangle}
&\stackrel{\textrm{GA}}{\to}\ 
\frac{1}{L^{2}}
\sideset{}{''}\sum\limits_{l_{\boldi},l_{\boldj},l_{\boldi}^{\prime},l_{\boldj}^{\prime}}
g_{l_{\boldi}}g_{l_{\boldj}}g_{l_{\boldi}^{\prime}}g_{l_{\boldj}^{\prime}} 
\langle  
\hat{c}_{\boldi a \sigma}^{\dagger}
\hat{c}_{\boldj b \sigma} 
\rangle_{0}
\dfrac{1}
{(1-n_{a}^{0})n_{b}^{0}}
\textstyle\prod\limits_{l}
\biggl(\dfrac{\bar{\Gamma}_{l}}{g_{l}^{2}}\biggr)^{-\Delta\bar{\Gamma}_{l}^{\prime\prime}}, 
\end{align}
respectively. 
Substituting Eq. (\ref{eq:X-g}) into these equations 
and using relations $L^{-\sum_{l} \Delta \bar{\Gamma}_{l}^{\prime}}=L$ 
and $L^{-\sum_{l} \Delta \bar{\Gamma}_{l}^{\prime\prime}}=L^{2}$ from 
Eqs. (\ref{eq:constraint-devi-int0}) and (\ref{eq:constraint-devi-kinetic0}), 
we find
\begin{align}
\dfrac{\langle \Psi_{\textrm{G}}| 
\hat{n}_{\boldi a \sigma}\hat{n}_{\boldi b \sigma^{\prime}} 
|\Psi_{\textrm{G}} \rangle}
{\langle \Psi_{\textrm{G}}  |\Psi_{\textrm{G}} \rangle}
&\stackrel{\textrm{GA}}{\to}\ 
\sideset{}{'}\sum\limits_{l_{\boldi}}
g_{l_{\boldi}}^{2} 
\exp\Bigl[-\sum_{l}(1+\lambda_{0})\Delta \bar{\Gamma}_{l}^{\prime}-\sum_{l,a^{\prime},\sigma^{\prime\prime}}\lambda_{a^{\prime}\sigma^{\prime\prime}}n_{la^{\prime}\sigma^{\prime\prime}}\Delta \bar{\Gamma}_{l}^{\prime}\Bigr]\notag\\
&= 
\sideset{}{'}\sum\limits_{l_{\boldi}}
g_{l_{\boldi}}^{2} 
\exp\Bigl[(1+\lambda_{0})+\sum_{a^{\prime},\sigma^{\prime\prime}}\lambda_{a^{\prime}\sigma^{\prime\prime}}n_{l_{\boldi}a^{\prime}\sigma^{\prime\prime}}\Bigr]\notag\\
&= 
\sideset{}{'}\sum\limits_{l_{\boldi}}
\dfrac{\bar{\Gamma}_{l_{\boldi}}}{L},\\[3pt]
\dfrac{\langle \Psi_{\textrm{G}}| 
\hat{c}_{\boldi a \sigma}^{\dagger}\hat{c}_{\boldj b \sigma}  
|\Psi_{\textrm{G}} \rangle}
{\langle \Psi_{\textrm{G}}  |\Psi_{\textrm{G}} \rangle}
&\stackrel{\textrm{GA}}{\to}\ 
\sideset{}{''}\sum\limits_{l_{\boldi},l_{\boldj},l_{\boldi}^{\prime},l_{\boldj}^{\prime}}
g_{l_{\boldi}}g_{l_{\boldj}}g_{l_{\boldi}^{\prime}}g_{l_{\boldj}^{\prime}} 
\langle 
\hat{c}_{\boldi a \sigma}^{\dagger}
\hat{c}_{\boldj b \sigma} 
\rangle_{0}
\dfrac{1}{(1-n_{a}^{0})n_{b}^{0}}
\exp\Bigl[-\sum_{l}(1+\lambda_{0})\Delta \bar{\Gamma}_{l}^{\prime\prime}\Bigr]
\exp\Bigl[-\sum_{l,a^{\prime},\sigma^{\prime\prime}}\lambda_{a^{\prime}\sigma^{\prime\prime}}n_{la^{\prime}\sigma^{\prime\prime}}\Delta \bar{\Gamma}_{l}^{\prime\prime}\Bigr]\notag\\
&= 
\sideset{}{''}\sum\limits_{l_{\boldi},l_{\boldj},l_{\boldi}^{\prime},l_{\boldj}^{\prime}}
g_{l_{\boldi}}g_{l_{\boldj}}g_{l_{\boldi}^{\prime}}g_{l_{\boldj}^{\prime}} 
\langle  
\hat{c}_{\boldi a \sigma}^{\dagger}
\hat{c}_{\boldj b \sigma} 
\rangle_{0}
\dfrac{1}{(1-n_{a}^{0})n_{b}^{0}}
\exp\Bigl[2(1+\lambda_{0})\Bigr]\notag\\
&\ \ \ \ \times 
\exp\Bigl[ \ \frac{1}{2}\sum_{a^{\prime},\sigma^{\prime}}\lambda_{a^{\prime}\sigma^{\prime}}(n_{l_{\boldi}a^{\prime}\sigma^{\prime}}+n_{l_{\boldj}a^{\prime}\sigma^{\prime}}+n_{l_{\boldi}^{\prime}a^{\prime}\sigma^{\prime}}+n_{l_{\boldj}^{\prime}a^{\prime}\sigma^{\prime}})\Bigr]\notag\\
&= 
\dfrac{1}{(1-n_{a}^{0})n_{b}^{0}}
\langle  
\hat{c}_{\boldi a \sigma}^{\dagger}
\hat{c}_{\boldj b \sigma} \rangle_{0}
\sideset{}{''}\sum\limits_{l_{\boldi},l_{\boldj},l_{\boldi}^{\prime},l_{\boldj}^{\prime}}
\dfrac{\sqrt{\bar{\Gamma}_{l_{\boldi}}\bar{\Gamma}_{l_{\boldj}}
\bar{\Gamma}_{l_{\boldi}^{\prime}}\bar{\Gamma}_{l_{\boldj}^{\prime}}}}
{L^{2}}.\label{eq:KE-general}
\end{align} 
In deriving Eq. (\ref{eq:KE-general}), 
we have also used the following relation 
due to Eq. (\ref{eq:Lag-sym2}): 
\begin{align}
\textstyle\sum\limits_{a=1,2}\lambda_{a \sigma}
(n_{l_{\boldi}^{\prime}a \sigma}+n_{l_{\boldj}^{\prime}a \sigma})
=
\textstyle\sum\limits_{a=1,2}\lambda_{a \sigma}
(n_{l_{\boldi}a \sigma}+n_{l_{\boldj}a \sigma}). 
\end{align}   

Finally, the variational energy becomes 
\begin{align}
E_{\textrm{gs}}
\stackrel{\textrm{GA}}{\to}
\min_{\{g_{l}\}} 
\biggl[&-
\sum\limits_{\boldi,\boldj}
\sum\limits_{a,b=1}^{2}
\sum\limits_{\sigma=\uparrow,\downarrow}
t_{ab}^{\boldi,\boldj}(\phi)
\dfrac{1}{(1-n_{1}^{0})n_{1}^{0}}
\langle \hat{c}_{\boldi a \sigma}^{\dagger}\hat{c}_{\boldj b \sigma} \rangle_{0}
\sideset{}{''}
\sum\limits_{l_{\boldi},l_{\boldj},l_{\boldi}^{\prime},l_{\boldj}^{\prime}}
\dfrac{\sqrt{\bar{\Gamma}_{l_{\boldi}}\bar{\Gamma}_{l_{\boldj}}
\bar{\Gamma}_{l_{\boldi}^{\prime}}\bar{\Gamma}_{l_{\boldj}^{\prime}}}}{L^{2}}
\notag\\
&-
\sum\limits_{\boldi,\boldj}
\sum\limits_{a,b=3}
\sum\limits_{\sigma=\uparrow,\downarrow}
t_{ab}^{\boldi,\boldj}(\phi)
\dfrac{1}{(1-n_{3}^{0})n_{3}^{0}}
\langle \hat{c}_{\boldi a \sigma}^{\dagger}\hat{c}_{\boldj b \sigma} \rangle_{0}
\sideset{}{''}\sum\limits_{l_{\boldi},l_{\boldj},l_{\boldi}^{\prime},l_{\boldj}^{\prime}}
\dfrac{\sqrt{\bar{\Gamma}_{l_{\boldi}}\bar{\Gamma}_{l_{\boldj}}
\bar{\Gamma}_{l_{\boldi}^{\prime}}\bar{\Gamma}_{l_{\boldj}^{\prime}}}}{L^{2}}
\notag\\
&+\ U
\sum\limits_{\boldi}
\sum\limits_{a=b}
\sum\limits_{\sigma=\uparrow}
\sum\limits_{\sigma^{\prime}=\downarrow}
\sideset{}{'}\sum\limits_{l_{\boldi}}
\dfrac{\bar{\Gamma}_{l_{\boldi}}}{L}
+ \ U^{\prime}
\sum\limits_{\boldi}
\sum\limits_{a>b}
\sum\limits_{\sigma,\sigma^{\prime}}
\sideset{}{'}\sum\limits_{l_{\boldi}}
\dfrac{\bar{\Gamma}_{l_{\boldi}}}{L}- \ J_{\textrm{H}}
\sum\limits_{\boldi}
\sum\limits_{a>b}
\sum\limits_{\sigma=\sigma^{\prime}}
\sideset{}{'}\sum\limits_{l_{\boldi}}
\dfrac{\bar{\Gamma}_{l_{\boldi}}}{L}
\biggr].\label{eq:GA-Egs-full}
\end{align}
Here, the restrictions about the summations 
with respect to $l_{\boldi}$, $l_{\boldj}$, $l_{\boldi}^{\prime}$, 
and $l_{\boldj}^{\prime}$ in the first and second terms 
are those for the hopping process 
$\hat{c}_{\boldi a \sigma}^{\dagger}\hat{c}_{\boldj b \sigma}$ ($a,b=1,2$) 
and hopping process $\hat{c}_{\boldi 3 \sigma}^{\dagger}\hat{c}_{\boldj 3 \sigma}$, 
and the restrictions about the summations 
with respect to $l_{\boldi}$ 
in the third, fourth, and final terms 
are those for the interaction terms $U$, $U^{\prime}$, and $J_{\textrm{H}}$.  
In deriving Eq. (\ref{eq:GA-Egs-full}), 
we have neglected the term of $\hat{H}_{\textrm{CEF}}$ in the variational energy 
since this term only gives a constant energy shift which is independent 
on the strength of the onsite interactions within the present treatment. 
Note that $E_{\textrm{gs}}$ can be written 
as the products of the component for orbital $a$ with spin $\sigma$ 
since we have neglected both 
$J^{\prime}$ and the transverse components of $J_{\textrm{H}}$ 
in the interacting Hamiltonian as denoted in Sec. II.  

\begin{table*}[tb]
\caption{The configurations, the wave functions, 
the values of $\hat{H}_{\textrm{int}}$, and the variational parameters $g_{X_{k}}$
for a PM state consisting of 
the degenerate $d_{xz}$ and $d_{yz}$ orbitals and the $d_{xy}$ orbital. 
$0$ and $\uparrow$ ($\downarrow$) in the wave functions 
mean there are no electrons in orbital $a$ with spin $\sigma$ 
and there is a spin-up (spin-down) electron in orbital $a$, 
respectively.}
\label{tab:GA}
\vspace{5pt}
\begin{tabular}{ccccccccc} \toprule\\[-10pt]
\multicolumn{1}{c}{ \ \ \ \ \ \ \ \ \ \ } &
\multicolumn{1}{c}{ \ \ \ \ Wave functions \ 
$|1\uparrow,1\downarrow;2\uparrow,2\downarrow;3\uparrow, 3\downarrow \rangle$ \ \ } & 
\multicolumn{1}{c}{ \ \ value of $\hat{H}_{\textrm{int}}$ \ \ } &
\multicolumn{1}{c}{ \ \ $g_{X_{k}}$ \ \ } \\[1pt] \hline\\[-8pt]
$X_{0}$
&$|0,0;0,0;0,0 \rangle$
&0&$g_{X_{0}}$\\[3pt]
$X_{1}$
&$|$$\uparrow,0;0,0;0,0 \rangle$, 
$|0,\downarrow;0,0;0,0 \rangle$, 
$|0,0;\uparrow,0;0,0 \rangle$, 
$|0,0;0,\downarrow;0,0 \rangle$
&0&$g_{X_{1}}$\\[3pt]
$X_{2}$
&$|0,0;0,0;\uparrow,0 \rangle$, 
$|0,0;0,0;0,\downarrow \rangle$
&0&$g_{X_{2}}$\\[3pt]
$X_{3}$
&$|$$\uparrow,\downarrow;0,0;0,0 \rangle$, 
$|0,0;\uparrow,\downarrow;0,0 \rangle$
&$U$&$g_{X_{3}}$\\[3pt]
$X_{4}$
&$|0,0;0,0;\uparrow,\downarrow \rangle$
&$U$&$g_{X_{4}}$\\[3pt]
$X_{5}$
&$|$$\uparrow,0;0,\downarrow;0,0 \rangle$, 
$|0,\downarrow;\uparrow,0;0,0 \rangle$, 
&$U^{\prime}$&$g_{X_{5}}$\\[3pt]
$X_{6}$
&$|$$\uparrow,0;0,0;0,\downarrow \rangle$, 
$|0,\downarrow;0,0;\uparrow,0 \rangle$, 
$|0,0;\uparrow,0;0,\downarrow \rangle$, 
$|0,0;0,\downarrow;\uparrow,0 \rangle$
&$U^{\prime}$&$g_{X_{6}}$\\[3pt]
$X_{7}$
&$|$$\uparrow,0;\uparrow,0;0,0 \rangle$, 
$|0,\downarrow;0,\downarrow;0,0 \rangle$
&$U^{\prime}-J_{\textrm{H}}$&$g_{X_{7}}$\\[3pt]
$X_{8}$
&$|$$\uparrow,0;0,0;\uparrow,0 \rangle$, 
$|0,\downarrow;0,0;0,\downarrow \rangle$, 
$|0,0;\uparrow,0;\uparrow,0 \rangle$, 
$|0,0;0,\downarrow;0,\downarrow \rangle$
&$U^{\prime}-J_{\textrm{H}}$&$g_{X_{8}}$\\[3pt]
$X_{9}$
&$|$$\uparrow,\downarrow;\uparrow,0;0,0 \rangle$, 
$|$$\uparrow,\downarrow;0,\downarrow;0,0 \rangle$, 
$|$$\uparrow,0;\uparrow,\downarrow;0,0 \rangle$, 
$|0,\downarrow;\uparrow,\downarrow;0,0 \rangle$
&$U+2U^{\prime}-J_{\textrm{H}}$&$g_{X_{9}}$\\[3pt]
$X_{10}$
&$|$$\uparrow,\downarrow;0,0;\uparrow,0 \rangle$, 
$|$$\uparrow,\downarrow;0,0;0,\downarrow \rangle$, 
$|0,0;\uparrow,\downarrow;\uparrow,0 \rangle$, 
$|0,0;\uparrow,\downarrow;0,\downarrow \rangle$
&$U+2U^{\prime}-J_{\textrm{H}}$&$g_{X_{10}}$\\[3pt]
$X_{11}$
&$|$$\uparrow,0;0,0;\uparrow,\downarrow \rangle$, 
$|0,\downarrow;0,0;\uparrow,\downarrow \rangle$, 
$|0,0;\uparrow,0;\uparrow,\downarrow \rangle$, 
$|0,0;0,\downarrow;\uparrow,\downarrow \rangle$
&$U+2U^{\prime}-J_{\textrm{H}}$&$g_{X_{11}}$\\[3pt]
$X_{12}$
&$|$$\uparrow,0;0,\downarrow;0,\downarrow \rangle$, 
$|0,\downarrow;\uparrow,0;\uparrow,0 \rangle$, 
$|$$\uparrow,0;0,\downarrow;\uparrow,0 \rangle$, 
$|0,\downarrow;\uparrow,0;0,\downarrow \rangle$, 
$|$$\uparrow,0;\uparrow,0;0,\downarrow \rangle$, 
$|0,\downarrow;0,\downarrow;\uparrow,0 \rangle$
&$3U^{\prime}-J_{\textrm{H}}$&$g_{X_{12}}$\\[3pt]
$X_{13}$
&$|$$\uparrow,0;\uparrow,0;\uparrow,0 \rangle$, 
$|0,\downarrow;0,\downarrow;0,\downarrow \rangle$
&$3U^{\prime}-3J_{\textrm{H}}$&$g_{X_{13}}$\\[3pt]
$X_{14}$
&$|$$\uparrow,\downarrow;\uparrow,0;0,\downarrow \rangle$, 
$|$$\uparrow,\downarrow;0,\downarrow;\uparrow,0 \rangle$, 
$|$$\uparrow,0;\uparrow,\downarrow;0,\downarrow \rangle$, 
$|0,\downarrow;\uparrow,\downarrow;\uparrow,0 \rangle$
&$U+5U^{\prime}-2J_{\textrm{H}}$&$g_{X_{14}}$\\[3pt]
$X_{15}$
&$|$$\uparrow,0;0,\downarrow;\uparrow,\downarrow \rangle$, 
$|0,\downarrow;\uparrow,0;\uparrow,\downarrow \rangle$
&$U+5U^{\prime}-2J_{\textrm{H}}$&$g_{X_{15}}$\\[3pt]
$X_{16}$
&$|$$\uparrow,\downarrow;\uparrow,0;\uparrow,0 \rangle$, 
$|$$\uparrow,\downarrow;0,\downarrow;0,\downarrow \rangle$, 
$|$$\uparrow,0;\uparrow,\downarrow;\uparrow,0 \rangle$, 
$|0,\downarrow;\uparrow,\downarrow;0,\downarrow \rangle$
&$U+5U^{\prime}-3J_{\textrm{H}}$&$g_{X_{16}}$\\[3pt]
$X_{17}$
&$|$$\uparrow,0;\uparrow,0;\uparrow,\downarrow \rangle$, 
$|0,\downarrow;0,\downarrow;\uparrow,\downarrow \rangle$
&$U+5U^{\prime}-3J_{\textrm{H}}$&$g_{X_{17}}$\\[3pt]
$X_{18}$
&$|$$\uparrow,\downarrow;\uparrow,\downarrow;0,0 \rangle$
&$2U+4U^{\prime}-2J_{\textrm{H}}$&$g_{X_{18}}$\\[3pt]
$X_{19}$
&$|$$\uparrow,\downarrow;0,0;\uparrow,\downarrow \rangle$, 
$|0,0;\uparrow,\downarrow;\uparrow,\downarrow \rangle$
&$2U+4U^{\prime}-2J_{\textrm{H}}$&$g_{X_{19}}$\\[3pt]
$X_{20}$
&$|$$\uparrow,0;\uparrow,\downarrow;\uparrow,\downarrow \rangle$, 
$|0,\downarrow;\uparrow,\downarrow;\uparrow,\downarrow \rangle$, 
$|$$\uparrow,\downarrow;\uparrow,0;\uparrow,\downarrow \rangle$,
$|$$\uparrow,\downarrow;0,\downarrow;\uparrow,\downarrow \rangle$
&$2U+8U^{\prime}-4J_{\textrm{H}}$&$g_{X_{20}}$\\[3pt]
$X_{21}$
&$|$$\uparrow,\downarrow;\uparrow,\downarrow;\uparrow,0 \rangle$, 
$|$$\uparrow,\downarrow;\uparrow,\downarrow;0,\downarrow \rangle$
&$2U+8U^{\prime}-4J_{\textrm{H}}$&$g_{X_{21}}$\\[3pt]
$X_{22}$
&$|$$\uparrow,\downarrow;\uparrow,\downarrow;\uparrow,\downarrow \rangle$
&$3U+12U^{\prime}-6J_{\textrm{H}}$&$g_{X_{22}}$\\[3pt] \toprule 
\end{tabular}
\end{table*}

For simplicity of the numerical calculation, 
we classify the possible 64 configurations 
into 23 groups in which the energies are same. 
These are shown in Table \ref{tab:GA}, where 
$X_{k}$ denotes the optimized number of sites with the configuration 
in the $k$ group; e.g., $X_{1}=\bar{\Gamma}_{l}$ for $l=1-4$, and so on. 
Denoting $x_{k}=X_{k}/L$, we obtain 
\begin{align}
E_{\textrm{gs}}  
\stackrel{\textrm{GA}}{\to}
\min_{x_{3}, x_{4}, \cdots ,x_{22}} 
\Bigl[
&-
\textstyle\sum\limits_{\boldi,\boldj}
\textstyle\sum\limits_{a,b=1}^{3}
\textstyle\sum\limits_{\sigma}
q_{ab}(x_{3},x_{4},\cdots, x_{22})
t_{ab}^{\boldi,\boldj}(\phi)
\langle \hat{c}^{\dagger}_{\boldi a \sigma} 
\hat{c}_{\boldj b \sigma} \rangle_{0}\notag\\
&+ U
\bigl(2x_{3}+x_{4}+4x_{9}+4x_{10}+4x_{11}
+4x_{14}+2x_{15}
+4x_{16}
+2x_{17}+2x_{18}+4x_{19}+8x_{20}\notag\\
&\ \ \ \ \ \ +4x_{21}
+3x_{22}\bigr)\notag\\
&+ U^{\prime}
\bigl(2x_{5}+4x_{6}+4x_{9}+4x_{10}
+4x_{11}+12x_{12}+12x_{14}
+6x_{15}+8x_{16}
+4x_{17}+2x_{18}
+4x_{19}\notag\\
&\ \ \ \ \ \ \ +16x_{20}
+8x_{21}+6x_{22}\bigr)\notag\\
&+ (U^{\prime}-J_{\textrm{H}})
\bigl(2x_{7}+4x_{8}+4x_{9}+4x_{10}+4x_{11}
+6x_{12}+6x_{13}+8x_{14}
+4x_{15}+12x_{16}+6x_{17}\notag\\
&\ \ \ \ \ \ \ \ \ \ \ \ \ \ \ \ \ +2x_{18}+4x_{19}+16x_{20}+8x_{21}+6x_{22}\bigr)\Bigr].
\label{eq:GA-Egs}
\end{align}
Here, $q_{ab}(x_{3},x_{4},\cdots, x_{22})$ is 
the RF of the kinetic energy for the Ru $t_{2g}$ orbital, which 
satisfies
\begin{align} 
q_{11}(x_{3},x_{4},\cdots, x_{22})=q_{12}(x_{3},x_{4},\cdots, x_{22})
=q_{21}(x_{3},x_{4},\cdots, x_{22})=q_{22}(x_{3},x_{4},\cdots, x_{22}).
\end{align} 
The RFs for the $d_{xz/yz}$ and $d_{xy}$ orbitals are given by 
\begin{align}
q_{11}(x_{3},x_{4},\cdots, x_{22})
=
\frac{1}{n_{1}^{0}(1-n_{1}^{0})}
\bigl[&\sqrt{x_{1}}
(\sqrt{x_{0}}+\sqrt{x_{3}}+\sqrt{x_{5}}+\sqrt{x_{7}}\ )
+\sqrt{x_{6}}
(\sqrt{x_{2}}+\sqrt{x_{10}}
+2\sqrt{x_{12}}\ )\notag\\
&+\sqrt{x_{8}}
(\sqrt{x_{2}}
+\sqrt{x_{10}}
+\sqrt{x_{12}}
+\sqrt{x_{13}}\ )
+\sqrt{x_{9}}
(\sqrt{x_{3}}+\sqrt{\bar{x}_{5}}
+\sqrt{x_{7}}+\sqrt{x_{18}}\ )\notag\\
&+\sqrt{x_{11}}
(\sqrt{x_{4}}+\sqrt{x_{15}}+\sqrt{x_{17}}+\sqrt{x_{19}}\ )
+\sqrt{x_{14}}
(\sqrt{x_{10}}+2\sqrt{x_{12}}+\sqrt{x_{21}}\ )\notag\\
&+\sqrt{x_{16}}
(\sqrt{x_{10}}+\sqrt{x_{12}}+\sqrt{x_{13}}+\sqrt{x_{21}}\ )
+\sqrt{x_{20}}
(\sqrt{x_{15}}+\sqrt{x_{17}}+\sqrt{x_{19}}+\sqrt{x_{22}}\ )\bigr]^{2},
\end{align}
and
\begin{align}
q_{33}(x_{3},x_{4},\cdots, x_{22})
=
\frac{1}{n_{3}^{0}(1-n_{3}^{0})}
\bigl[
&\sqrt{x_{2}}
(\sqrt{x_{0}}
+\sqrt{x_{4}}\ )
+2\sqrt{x_{1}}
(\sqrt{x_{6}}
+\sqrt{x_{8}}\ )
+2\sqrt{x_{10}}
(\sqrt{x_{3}}+\sqrt{x_{19}}\ )\notag\\
&+2\sqrt{x_{11}}
(\sqrt{x_{6}}
+\sqrt{x_{8}}\ )
+2\sqrt{x_{12}}
(\sqrt{x_{5}}+\sqrt{x_{15}}\ )
+\sqrt{x_{17}}
(\sqrt{x_{12}}
+\sqrt{x_{13}}\ )\notag\\
&+\sqrt{x_{7}}
(\sqrt{x_{12}}
+\sqrt{x_{13}}\ )
+2\sqrt{x_{9}}
(\sqrt{x_{14}}
+\sqrt{x_{16}}\ )
+2\sqrt{x_{20}}
(\sqrt{x_{14}}
+\sqrt{x_{16}}\ )\notag\\
&+\sqrt{x_{21}}
(\sqrt{x_{18}}+\sqrt{x_{22}}\ )\bigr]^{2},
\end{align}
respectively. 
Note that the optimization with respect to $\{g_{X_{k}}\}$ 
is equivalent that with respect to $\{x_{k}\}$ 
due to the relation $x_{k}=g_{X_{k}}^{2}
\exp(1+\lambda_{0}+\sum_{a, \sigma}\lambda_{a\sigma}n_{la\sigma})$. 
In the numerical calculations, 
we use the following constraints 
instead of determining the Lagrange multipliers $\lambda_{0}$ 
and $\lambda_{a\sigma}$:
\begin{align}
1
=&\ x_{0}
+4x_{1}+2x_{2}
+2x_{3}+x_{4}+2x_{5}+4x_{6}+2x_{7}
+4x_{8}+4x_{9}+4x_{10}
+4x_{11}+6x_{12}+2x_{13}+4x_{14}
+2x_{15}\notag\\
&+4x_{16}
+2x_{17}+x_{18}+2x_{19}
+4x_{20}+2x_{21}+x_{22}, 
\label{eq:constraint-x0}\\
n_{1}^{0}
=&\  
x_{1}
+x_{3}+x_{5}+x_{6}+x_{7}+x_{8}
+3x_{9}+2x_{10}+x_{11}
+3x_{12}+x_{13}+3x_{14}+x_{15}+3x_{16}+x_{17}
+x_{18}+x_{19}\notag\\
&+3x_{20}+2x_{21}+x_{22}, \label{eq:constraint-x1}\\
n_{3}^{0}
=\ & 
x_{2}
+x_{4}+2x_{6}+2x_{8}+2x_{10}+4x_{11}
+3x_{12}+x_{13}+2x_{14}+2x_{15}+2x_{16}
+2x_{17}+2x_{19}+4x_{20}+x_{21}+x_{22}.\label{eq:constraint-x2}
\end{align}
\end{widetext}

\section{Results}
In this section, 
we show the numerical results of the GA 
for three cases with the effective models of $x=2$ and $0.5$ 
and the special model. 
The variational energy Eq. (\ref{eq:GA-Egs}) is numerically minimized 
with respect to $\{x_{k}\}$ under the constraints 
(\ref{eq:constraint-x0}){--}(\ref{eq:constraint-x2}) 
by Powell's method,~\cite{Numerical recipe} 
which is one of the numerical methods to minimize a function 
with more than one variable. 
In this work, 
we use the value of $U$ as the parameter, and 
set $U^{\prime}=U-2J_{\textrm{H}}$, $J_{\textrm{H}}=U/4$, 
and $W_{\textrm{tot}}=4.1$ eV, which is obtained in the effective model of $x=2$.   

\subsection{Mass enhancement for the effective models of $x=2$ and $0.5$}
\begin{figure}[tb]
\vspace{-8pt}
\includegraphics[width=86mm]{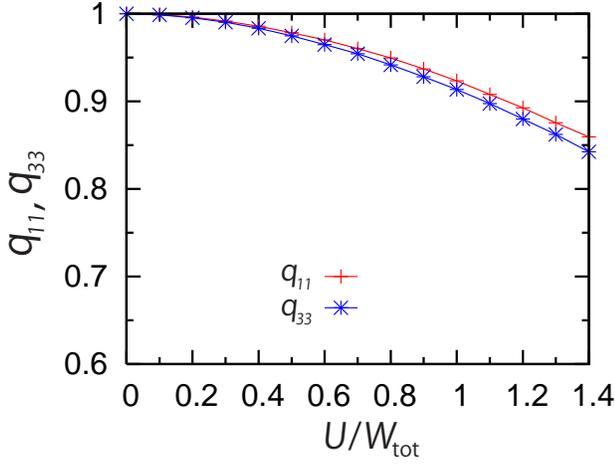}
\vspace{-20pt}
\caption{(Color online) 
RFs for the $d_{xz/yz}$ and $d_{xy}$ orbitals 
as a function of $U/W_{\textrm{tot}}$ for the effective model of $x=2$. 
We set $J_{\textrm{H}}=U/4$. 
}
\label{fig:renorm0-JH}
\end{figure}
\begin{figure}[tb]
\vspace{4pt}
\includegraphics[width=86mm]{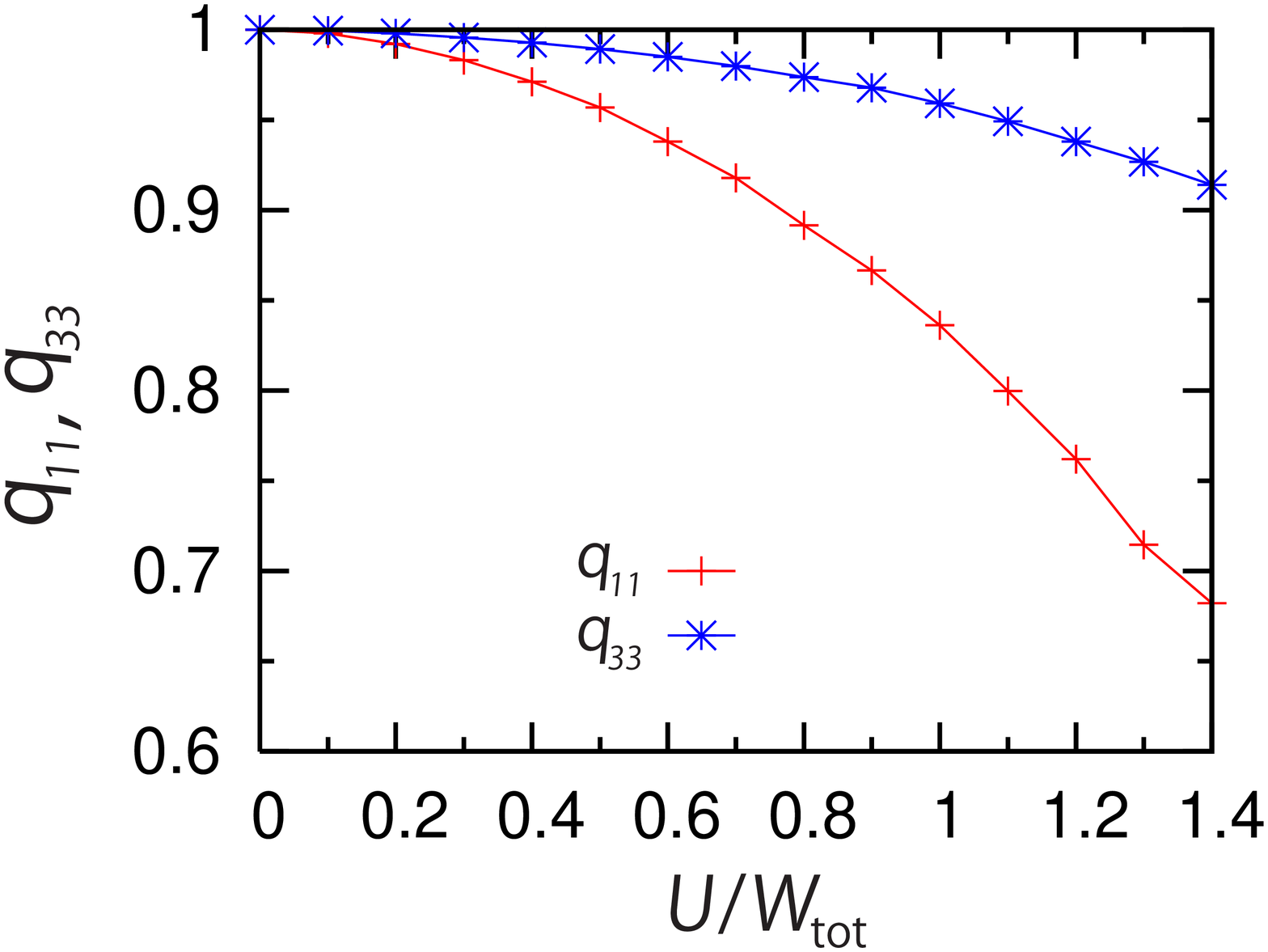}
\vspace{-20pt}
\caption{(Color online) 
RFs for the $d_{xz/yz}$ and $d_{xy}$ orbitals 
as a function of $U/W_{\textrm{tot}}$ for the effective model of $x=0.5$. 
We set $J_{\textrm{H}}=U/4$. }
\label{fig:renorm20-05}
\end{figure}
\begin{figure}[tb]
\vspace{-8pt}
\includegraphics[width=86mm]{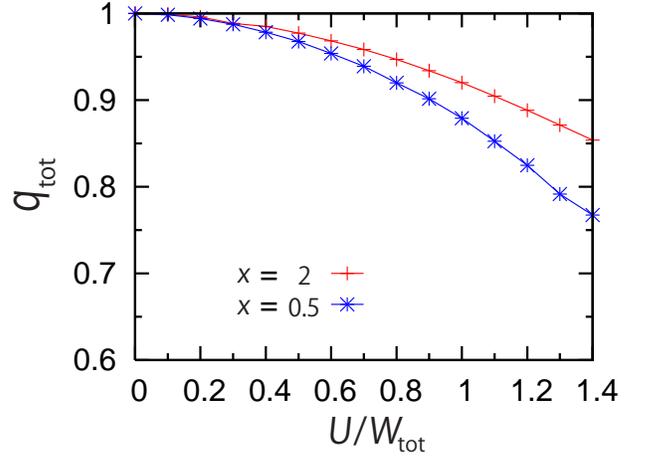}
\vspace{-20pt}
\caption{(Color online) 
Total RFs as a function of $U/W_{\textrm{tot}}$ 
for the effective models of $x=2$ and $0.5$. 
We set $J_{\textrm{H}}=U/4$. }
\label{fig:renorm-norm}
\end{figure}

We first show the results for the effective models of $x=2$ and $0.5$. 
Figures \ref{fig:renorm0-JH} and \ref{fig:renorm20-05} represent 
the RFs of the kinetic energy for the Ru $t_{2g}$ orbitals 
as a function of $U/W_{\textrm{tot}}$ for these models.  
We see from Fig. \ref{fig:renorm0-JH} that 
the RFs for the $d_{xz/yz}$ and $d_{xy}$ orbitals 
are nearly the same for the case of $x=2$. 
This is probably originated from the fact that 
these orbitals have nearly the same occupation numbers 
(i.e., $n_{1}^{0}=n_{2}^{0}=n_{3}^{0}=4/3$) in the absence of the interactions. 
On the other hands, for the case of $x=0.5$, 
we find from Fig. \ref{fig:renorm20-05} that 
the RFs for the $d_{xz/yz}$ and $d_{xy}$ orbitals are different, 
and that the increase of $U/W_{\textrm{tot}}$ leads to 
the large difference between these RFs. 
The occupation number for each Ru $t_{2g}$ orbital becomes 
$(n_{1}^{0},n_{2}^{0},n_{3}^{0})=(1.17,1.17,1.66)$ for the case of $x=0.5$. 
This change of the occupation numbers results mainly from 
the downward shift of the $d_{xy}$ orbital 
since we have $(n_{1}^{0},n_{2}^{0},n_{3}^{0})=(1.32,1.32,1.35)$, 
which are little different from those for the case of $x=2$, 
for the model setting $\phi=15^{\circ}$ 
and $\Delta_{t_{2g}}=0$ eV in Eq. (\ref{eq:H-phi}). 
By using the analogy with the result 
for the single-orbital Hubbard model, 
the difference of the RFs between the cases of $x=2$ and $x=0.5$ 
will be due to this change of the occupation numbers 
approaching the integer values towards $x=0.5$, 
which is expected in the usual Mott transition; 
in the present case, the occupation numbers expected 
in the Mott insulator 
are $1$ for the $d_{xz/yz}$ orbital and $2$ for the $d_{xy}$ orbital, 
respectively. 

In order to compare our results 
with the experimentally observed mass enhancement, 
we define a total RF, 
\begin{align} 
q_{\textrm{tot}}= \sqrt{\frac{1}{3}
\textstyle\sum\limits_{a=1}^{3}q_{aa}^{2}},
\end{align} 
which estimates the inverse of the mass enhancement. 
Figure \ref{fig:renorm-norm} shows the total RFs 
as a function of $U/W_{\textrm{tot}}$ 
for the effective models of $x=2$ and $0.5$.  
We see that 
$q_{\textrm{tot}}$ for the case of $x=0.5$ becomes much smaller 
than that for the case of $x=2$ 
as $U/W_{\textrm{tot}}$ increases. 
Therefore, our results suggest that 
moderately strong Coulomb interaction and 
the modifications of the electronic structures 
for the Ru $t_{2g}$ orbitals 
due to the rotation of RuO$_{6}$ octahedra lead to 
mass enhancement for $x=0.5$ than that for $x=2$. 

\subsection{Primary effect of the Ca substitution \\on mass enhancement 
around $x=0.5$}
\begin{figure}[tb]
\vspace{14pt}
\includegraphics[width=86mm]{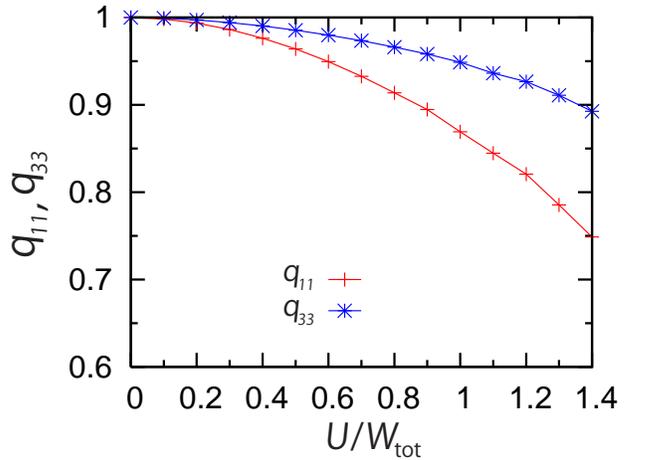}
\vspace{-20pt}
\caption{(Color online) 
RFs for the $d_{xz/yz}$ and $d_{xy}$ orbitals 
as a function of $U/W_{\textrm{tot}}$ 
for the special model. 
We set $J_{\textrm{H}}=U/4$. 
}
\label{fig:renorm-vHs}
\end{figure}
\begin{figure}[tb]
\vspace{-8pt}
\includegraphics[width=86mm]{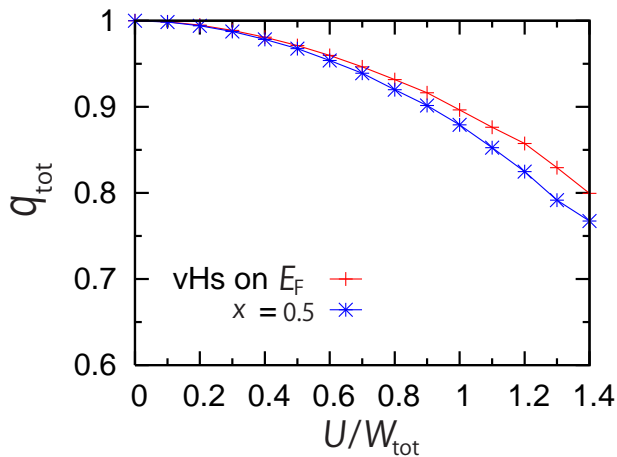}
\vspace{-20pt}
\caption{(Color online) 
Total RFs as a function of $U/W_{\textrm{tot}}$ 
for the effective model of $x=0.5$ and 
the special model. 
We set $J_{\textrm{H}}=U/4$. 
}
\label{fig:renorm-compare}
\end{figure}

In order to clarify the role of the Ca substitution in forming the HF, 
we calculate the RF for the special model. 
Figure \ref{fig:renorm-vHs} shows the RFs for each Ru $t_{2g}$ orbital. 
We see that the RF for the $d_{xz/yz}$ orbital is slightly larger
than that for the case of $x=0.5$, 
while the RF for the $d_{xy}$ orbital is slightly smaller. 
The former results mainly from the decrease of the occupation number 
for the $d_{xz/yz}$ orbital from $1.21$ to $1.17$, 
and the latter results mainly from the increase of the occupation number 
for the $d_{xy}$ orbital from $1.57$ to $1.66$; 
both changes of the occupation numbers arise from 
the downward shift for the $d_{xy}$ orbital, 
which is induced by the rotation of RuO$_{6}$ octahedra. 

Figure \ref{fig:renorm-compare} shows 
the total RFs for both the effective model of $x=0.5$ and the special model. 
We find that 
the inverse of $q_{\textrm{tot}}$ for the effective model of $x=0.5$ 
is larger than that for the special model, 
and that Coulomb interaction enhances 
this difference of the inverse of $q_{\textrm{tot}}$. 
Therefore, our result indicates that  
the vHs for the $d_{xy}$ orbital plays a secondary role 
in enhancing the effective mass around $x=0.5$; 
the primary one arises from 
the change of the occupation numbers 
approaching the integer values. 
Note that these occupation numbers are different from 
those expected in the OSMT 
but same to those expected in the usual Mott transition, 
where the occupation numbers for the $d_{xz/yz}$ and $d_{xy}$ orbitals 
are 1 and 2, respectively.   

\section{Discussion}
\subsection{Comparison with previous theoretical studies}
We first remark on the physical meaning of the enhancement of 
the effective mass obtained in the GA. 
For simplicity, we consider a PM state consisting of a single orbital. 
The following argument is based on the discussion in Ref. 33. 
One of the origins of the HF behavior 
is the criticality approaching the Mott transition. 
Namely, 
the formation of the HF results from 
separation of energy scale between spin and charge 
degrees of freedom due to electron correlation; 
on-site charge fluctuation suppresses at the higher scale 
corresponding to $U$, 
and the lower one, $T_{\textrm{F}}^{\ast}$, associates with 
the local spin fluctuation. 
Below $T_{\textrm{F}}^{\ast}$, a Fermi-liquid description applies; 
the mass enhancement is inversely proportional 
to a ratio of $T_{\textrm{F}}^{\ast}$ to the noninteracting Fermi temperature. 
Although the analysis using the GA restricts 
to the discussion about the properties for the ground states 
and can not address the possibility of the above separation of energy scale, 
the GA can analyze the effective mass for systems with a different parameter. 
A similar argument can apply to a system with orbital degrees of freedom. 
Therefore, we think that the analysis using the GA can capture 
the tendency of mass enhancement in Ca$_{2-x}$Sr$_{x}$RuO$_{4}$ 
for $0.5\leq x\leq 2$.

In the present calculation, 
we have not discussed Mott transition. 
From a theoretical point of view, 
the previous DMFT caluculation~\cite{Liebsch} based on 
the models for Ca$_{2-x}$Sr$_{x}$RuO$_{4}$ in $0.5\leq x \leq 2$ 
has shown that 
the critical value of $U$ for the Mott transition, $U_{\textrm{c}}$,
at $x=0.5$ is larger than 6 eV (i.e., $U/W_{\textrm{tot}}> 1.4$) at $T=0.02$ eV.
In principle, the Gutzwiller-type variational wave function 
gives the Mott transition as a function of $U$, 
when charge fluctuations, which play an important role 
in the vicinity of the Mott transition,~\cite{Furukawa-Imada} 
are included.~\cite{VMC-Yokoyama,VMC1,VMC2,VMC3,VMC4} 
However, the present study using the GA is 
inappropriate to address the possibility of the Mott transition. 
Thus, we have shown the results 
only for the case with $U/W_{\textrm{tot}}\leq 1.4$ 
where the system will remain metallic. 

The present formalism of the GA neglects 
the variation of the occupation number for each orbital 
due to electron correlation for simplicity.  
The previous DMFT calculation~\cite{Liebsch} based on 
the models for Ca$_{2-x}$Sr$_{x}$RuO$_{4}$ with $0.5\leq x\leq 2$ 
has observed a drastic variation of the occupation numbers 
($\sim 20-30\%$) for values of $U$ near $U_{\textrm{c}}$, 
while the variation is about $10\%$ (a few $\%$) 
for $U\sim 0.8U_{\textrm{c}}$ ($U\sim 0.7U_{\textrm{c}}$).  
This indicates that 
the variation of the occupation numbers due to electron correlation 
is important near the Mott transition. 
In this paper, 
we have discussed the HF behavior 
only in the regions where 
the effects of the variation of the occupation numbers is small. 

Although it will be necessary to include 
the variation of the occupation number for each orbital 
for the quantitative argument, 
we think from the following arguments 
that the obtained tendency of the mass enhancement in $0.5\leq x\leq 2$ 
does not change qualitatively 
even if the variation of the occupation number for each orbital is included. 
According to the DMFT study~\cite{Liebsch} 
for the three-orbital Hubbard model, 
$U_{\textrm{c}}$ decreases 
when the occupation number for each orbital in the absence of interactions 
approaches to an integer value. 
Simultaneously, the occupation number for each orbital changes 
as a function of $U/W_{\textrm{tot}}$. 
Since the decrease of $U_{\textrm{c}}$ leads to 
the increase of the effective mass 
at a fixed value of $U/W_{\textrm{tot}}$ and $J_{\textrm{H}}/U$, 
the mass enhancement will be largest at $x=0.5$ 
where the occupation numbers in the absence of interactions are 
nearest to integer values in $0.5\leq x \leq 2$. 
Therefore, 
our results about the mass enhancement 
in $0.5\leq x\leq 2$ will not change qualitatively 
even if the variation of the occupation number for each orbital is included.   

DMFT studies~\cite{Haule-FePn,Haule-Ru,Georges-Hund} proposed that 
the Hund's rule coupling plays an important role 
in stabilizing a metallic state with large effective mass. 
For example, 
a DMFT study~\cite{Georges-Hund} 
for the degenerate three-orbital Hubbard model 
showed that 
the increase of $J_{\textrm{H}}/U$ leads to 
a decrease of $U_{\textrm{c}}$ at $\frac{1}{2}$-filling and 
an increase of $U_{\textrm{c}}$ at $\frac{1}{6}$-filling. 
(Note that there are three electrons per a site at $\frac{1}{2}$-filling 
and there is one electron per a site at $\frac{1}{6}$-filling.) 
From this result, they claimed that 
the increase of $J_{\textrm{H}}/U$ extends 
a region for the metallic state with large effective mass 
at the filling away from $\frac{1}{2}$-filling; 
this metallic state is called Hund's metal.  
However, in Ca$_{2-x}$Sr$_{x}$RuO$_{4}$, 
the total filling number does not change 
in $0.5\leq x \leq 2$ (i.e., $\frac{2}{3}$-filling) and 
the value of $J_{\textrm{H}}/U$ does not change either. 
The main changes due to the Ca substitution are 
both the bandwidth for each Ru $t_{2g}$ orbital 
and the occupation number for each Ru $t_{2g}$ orbital. 
Therefore, 
the effect of the Hund's rule coupling is not important 
for the difference between $x=0.5$ and $2$. 
Instead, 
the criticality approaching the Mott transition plays a more important role 
in enhancing the effective mass in Ca$_{2-x}$Sr$_{x}$RuO$_{4}$ towards $x=0.5$. 

We now address the effect of the neglected terms in 
the interacting Hamiltonian (i.e., $J^{\prime}$ and 
the transverse components of $J_{\textrm{H}}$) on the electronic states. 
According to the previous study~\cite{GA-multi} of the PM state 
for the two-orbital Hubbard model in the GA, 
these terms slightly stabilize a PM metal against a nonmagnetic insulator.
This result suggests that 
the effect of these neglected terms little affects 
the properties for the PM state studied in the present study within the GA. 
On the other hand, the transverse components of $J_{\textrm{H}}$ 
will strongly affect the stability of the magnetically ordered states. 
Therefore, 
these neglected terms will not change 
the obtained tendency of the mass enhancement in PM states. 

According to the density-functional calculation 
for Ca$_{2-x}$Sr$_{x}$RuO$_{4}$ in $0.5\leq x \leq 2$ 
within the LDA,~\cite{Oguchi-LS} 
the total bandwidth for the Ru $t_{2g}$ orbitals 
becomes smaller for $x=0.5$ than that for $x=2$. 
Although this reduction is partially included in our calculation 
through the $\phi$ dependence of the dispersions, 
it seems that these are other effects which 
reduce the total band width. 
Thus, if this effect is fully included, 
the difference between the values of $q_{\textrm{tot}}^{-1}$ 
for $x=2$ and $0.5$
will become larger than for the present calculation. 

In our analysis, 
we have assumed that 
the roles of the O $2p$ orbitals are only to change the $dp$ hybridizations. 
However, the previous theoretical work~\cite{Yoshioka} 
based on the $dp$ model for Sr$_{2}$RuO$_{4}$ 
has proposed that 
Coulomb interaction for the O $2p$ orbitals plays an important role 
in stabilizing spin-triplet superconductivity. 
We expect that the Coulomb interaction for the O $2p$ orbitals 
leads to a larger mass enhancement than the present calculation. 
The more detailed theoretical study about 
the role of the O $2p$ orbitals is deserved. 

There is a theoretical proposal that 
the vHs for the $d_{xy}$ orbital plays a primary role 
in forming HFs around $x=0.5$ 
on the basis of the density-functional calculation 
within the LDA.~\cite{hyb-t2g-eg} 
This work has proposed that 
the rotation-induced hybridization of the $d_{xy}$ orbital 
to the $d_{x^{2}\textrm{-}y^{2}}$ orbital causes 
the magnetic instability due to the nesting of the FS 
for the $d_{xy}$ orbital, 
and that the instability will lead to the mass enhancement around $x=0.5$. 
In contrast, 
our results suggest that 
the primary role arises from the criticality approaching 
the usual Mott transition 
resulting from the change of the occupation numbers 
for the Ru $t_{2g}$ orbitals 
due to the downward shift of the $d_{xy}$ orbital, 
and not from the vHs. 
It will be necessary to study the effect of the vHs 
on the formation of HFs around $x=0.5$ more systematically. 

\subsection{Correspondence with experimental results} 
We first discuss the roles of the Ru $e_{g}$ and O $2p$ orbitals 
in determining the electronic states 
for Ca$_{2-x}$Sr$_{x}$RuO$_{4}$ in $0.5\leq x \leq 2$. 
In this study, 
we have taken account of the effects of these orbitals 
as the changes of the $dp$ hybridization 
and the CEF energy for the Ru $t_{2g}$ orbitals; 
the latter is 
the downward shift of the $d_{xy}$ orbital 
due to the hybridization with the $d_{x^{2}\textrm{-}y^{2}}$ orbital. 
There is no experimental evidence that 
the Ru $e_{g}$ orbitals play an important role 
in determining the electronic states 
except the possible change of the CEF energy 
through the hybridization of the Ru $t_{2g}$ orbitals; 
thus, 
our treatment about the Ru $e_{g}$ orbitals will be valid. 
In contrast, a polarized neutron diffraction measurement 
for Ca$_{1.5}$Sr$_{0.5}$RuO$_{4}$ 
has observed a field-induced magnetic moment on the in-plane O ions, 
which is about 20$\%$ of that for Ru ions.~\cite{polarized-neutron} 
This result indicates 
that not only the $dp$ hybridizations, 
but also the Coulomb interaction for the O $2p$ orbitals will play 
non-negligible roles in determining the electronic states. 
It is thus necessary to study the role of the O $2p$ orbitals systematically.  

Let us remark on the role of the spin-orbit interaction, 
which has been neglected in this work. 
The experimentally observed FSs~\cite{dHvA,ARPES05} for $x=2$ and $0.5$ 
are reproducible by the density-function calculations 
without the spin-orbit interaction.~\cite{Oguchi,Singh,Oguchi-LS} 
Therefore, the spin-orbit interaction will be negligible 
in determining the electronic states. 

We have also neglected 
the effect of the disorder induced by the Ca substitution in this work. 
A measurement with a SQUID magnetometer 
for Ca$_{1.5}$Sr$_{0.5}$RuO$_{4}$ has observed 
glassy behavior (i.e., the time-dependent magnetization),~\cite{Nakatsuji-HF} 
which is similar to the behavior observed 
in Ca$_{0.95}$Sr$_{0.05}$RuO$_{3}$.~\cite{CSRO3-FM} 
This behavior is related to the disorder by the Ca substitution. 
It is thus necessary to include 
the effect of the disorder on the electronic state 
in order to discuss the electronic states for Ca$_{2-x}$Sr$_{x}$RuO$_{4}$. 
This remains as a future problem. 

Let us discuss the role of the vHs for the $d_{xy}$ orbital. 
Experimentally, 
substitution of La$^{3+}$ for Sr$^{2+}$ in Sr$_{2}$RuO$_{4}$ leads to 
the downward shift of the vHs towards the Fermi level 
without any structural distortions.~\cite{rigid-band-La214} 
In this case, 
the coefficient of the electronic specific heat for Sr$_{1.8}$La$_{0.2}$RuO$_{4}$ 
reaches $1.3$ times of that in Sr$_{2}$RuO$_{4}$.~\cite{La-214} 
This mass enhancement is mainly due to 
the increase of the DOS by the vHs for the $d_{xy}$ orbital;  
the density-functional calculation within the LDA 
and the ARPES measurement 
support this mechanism.~\cite{rigid-band-La214,ARPES-La214} 
However, 
in the case of Ca$_{2-x}$Sr$_{x}$RuO$_{4}$ with $x=0.5$,  
the ARPES measurement~\cite{ARPES05} has shown that 
the vHs is located below the Fermi level. 
Therefore, 
there must be other factors other than the vHs in enhancing the effective mass 
in Ca$_{2-x}$Sr$_{x}$RuO$_{4}$; 
our results indicate that 
one of the factors is the criticality approaching 
the usual Mott transition. 
 
We next remark on the roles of the spin fluctuations. 
Around $x=0.5$, 
the resistivity behaves metallic (i.e., $d\rho/dT>0$), and 
the spin susceptibility shows Curie-Weiss behavior, as described in Sec. I. 
These experimental facts indicate that 
the mode-mode coupling for the spin fluctuations plays an important role 
in the electronic states around $x=0.5$ 
since the enhancement of the mode-mode coupling generally 
leads to the Curie-Weiss behavior.~\cite{Moriya-SCR} 
The study taking account of the mode-mode coupling is 
a remaining future problem. 

There are several experimental results which indicate 
that ferromagnetic spin fluctuation also plays an important role in 
the electronic states around $x=0.5$. 
One of the examples is the enhancement of the Wilson ration 
towards $x=0.5$.~\cite{Nakatsuji-lattice} 
This result indicates that 
the system is a nearly ferromagnetic metal 
near $x=0.5$.~\cite{Berk-Schrieffer,NearlyFM} 
In addition, 
the inelastic neutron scattering measurement~\cite{neutron-gamma} 
has claimed that 
the value of $\gamma_{\textrm{e}}$ for $x=0.62$ is reproducible 
by a phenomenological theory,~\cite{Moriya-gamma} 
including the over-damped magnetic excitations 
for the ferromagnetic fluctuation. 
However, 
the additional Ca substitution does not lead to ferromagnetism 
but evolves the short-range antiferromagnetic correlation in $0.2\leq x< 0.5$ 
and the long-range antiferromagnetic correlation 
in $0\leq x<0.2$.~\cite{Nakatsuji-lattice} 
This experimental fact suggests that 
Ca$_{2-x}$Sr$_{x}$RuO$_{4}$ around $x=0.5$ 
can not be regarded as a simple nearly ferromagnetic metal. 
Furthermore, 
the value of $\gamma_{\textrm{e}}$ for $x=2$ 
is insensitive to the magnetic fields up to $14$ T, 
indicating that the mass enhancement for $x=2$ 
is not due to spin fluctuations.~\cite{Maeno-RW} 
It is thus needed to study the role of the ferromagnetic spin fluctuation.  

\section{Summary}
In order to clarify the origin of the HF behavior around $x=0.5$, 
we have studied the electronic states for Ca$_{2-x}$Sr$_{x}$RuO$_{4}$ 
in $0.5\leq x \leq 2$ within the GA 
on the basis of the three-orbital Hubbard model for the Ru $t_{2g}$ orbitals. 
We have assumed that 
the Ca substitution affects the electronic structures mainly by 
the changes of the $dp$ hybridizations 
between the Ru $4d$ and O $2p$ orbitals, 
and have estimated the mass enhancement 
on the basis of the models taking account of 
these effects on the electronic structures. 
In particular, 
we have numerically calculated the RF within the GA 
for three cases with the effective models of $x=2$ and $0.5$ 
and the special model. 
We have found that 
the inverse of the total RF becomes the largest for the case of $x=0.5$, 
and that the vHs for the $d_{xy}$ orbital plays a secondary role 
in enhancing the effective mass. 
Our results can reproduce the experimentally observed 
tendency of the effective mass in $0.5\leq x \leq 2$:~\cite{Nakatsuji-HF} 
the coefficient of the electronic specific heat 
monotonically increases towards $x=0.5$ 
although the vHs appears on the Fermi level at smaller Sr concentration 
than $x=0.5$. 

Our calculation suggests that 
the HF behavior around $x=0.5$ comes from 
the cooperative effects between 
moderately strong Coulomb interaction compared to 
the total bandwidth 
and the modification of the electronic structures 
due to the rotation of RuO$_{6}$ octahedra 
(i.e., the variation of the $dp\pi$ hybridizations and 
the downward shift for the $d_{xy}$ orbital). 
We propose that 
moderately strong electron correlation and 
the orbital-dependent modifications of the electronic structures 
due to the lattice distortions 
play important roles in the electronic states for Ca$_{2-x}$Sr$_{x}$RuO$_{4}$.

\begin{acknowledgments}
The authors would like to thank Y. Yanase, H. Watanabe, 
and T. Kariyado for useful comments. 
This work is supported by a Grant-in-Aid for 
Scientific Research on Innovative Areas 
``Heavy Electrons'' (No. 20102008) of The Ministry of Education, 
Culture, Sports, Science, and Technology, Japan. 
\end{acknowledgments}


\end{document}